\documentclass[superscriptaddress,english,floatfix,twocolumn,showpacs,aps,reprint, prb]{revtex4-2}

\usepackage{amsmath, amssymb, amsfonts}
\usepackage{graphicx}
\usepackage{bm}
\usepackage{dsfont}
\usepackage{xfrac}
\usepackage{siunitx}
\usepackage{nicefrac} 
\usepackage{hyperref}
\usepackage[normalem]{ulem}
\usepackage{xcolor}
\usepackage{booktabs}

\hypersetup{colorlinks=true, citecolor=blue, urlcolor=blue, linkcolor=blue, breaklinks=true}
\definecolor{go_green}{rgb}{0.13, 0.55, 0.13}

\begin{document}

\title{Three-state Potts nematic order in stacked frustrated spin models with SO(3) symmetry 
}

\author{Ana-Marija Nedi\' c}
\author{Victor L. Quito}
\author{Yuriy Sizyuk}
\author{Peter P. Orth}
\affiliation{Department of Physics and Astronomy, Iowa State University, Ames, Iowa 50011, USA}
\affiliation{Ames National Laboratory, Ames, Iowa 50011, USA}

\date{\today}

\begin{abstract}
We propose stacked two-dimensional lattice designs of frustrated and SO(3) symmetric spin models consisting of antiferromagnetic (AFM) triangular and ferromagnetic (FM) sixfold symmetric sublattices that realize emergent $\mathbb{Z}_3$ Potts nematic order. Considering bilinear-biquadratic spin interactions, our models describe an SO(3)-symmetric triangular lattice AFM subject to a fluctuating magnetization arising from the FM coupled sublattice. 
We focus on the classical AFM-FM windmill model and map out the zero- and finite-temperature phase diagram using Monte Carlo simulations and analytical calculations. We discover a state with composite Potts nematic order above the ferrimagnetic three-sublattice up-up-down ground state and relate it to Potts phases in SO(3)-broken Heisenberg and Ising AFMs in external magnetic fields. Finally, we show that the biquadratic exchange in our model is automatically induced by thermal and quantum fluctuations in the purely bilinear Heisenberg model, easing the requirements for realizing these lattice designs experimentally.
\end{abstract}
\maketitle

\section{Introduction}
\label{Sec:Introduction}
Frustrated triangular-lattice Heisenberg and Ising antiferromagnets that are subject to external magnetic fields exhibit rich phase diagrams, including finite-temperature $\mathbb{Z}_3$ Potts phase transitions into magnetically ordered states~\cite{Starykh2015, Gvozdikova2011, Seabra2011, alexanderLatticeGasTransition1975, Metcalf1973, Schick1976}. The classical triangular Heisenberg antiferromagnet in a magnetic field, which explicitly breaks SO(3) symmetry, exhibits an extensive and continuous ground-state degeneracy that is lifted by thermal~\cite{Kawamura1984} and quantum fluctuations~\cite{Chubukov1991} via an order-by-disorder mechanism~\cite{Villain1977, Villain1980, shenderAntiferromagneticGarnetsFluctuationally1982,Henley_PRL_1989}. In an intermediate field range, a collinear three-sublattice up-up-down (UUD) state is selected that spontaneously breaks $\mathbb{Z}_3$ symmetry (corresponding to the position of the minority down spin). This results in the stabilization of a one-third magnetization plateau both at zero and finite temperatures $T$. The magnetic transition at finite $T$ lies in the $\mathbb{Z}_3$ Potts universality class~\cite{Seabra2011}. Similarly, the triangular Ising AFM in a magnetic field develops long-range UUD magnetic order at finite temperature via a Potts phase transition~\cite{Metcalf1973, kinzelPhenomenologicalScalingApproach1981, qianCriticalFrontierTriangular2004}. At $T=0$ the transition requires a nonzero critical field value and lies in the Kosterlitz-Thouless (KT) universality class~\cite{nienhuisTriangularSOSModels1984,qianCriticalFrontierTriangular2004}.  

Two-dimensional (2D) spin models with continuous SO(3) symmetry, on the other hand, cannot develop long-range magnetic order at $T>0$ due to the Hohenberg-Mermin-Wagner theorem~\cite{Hohenberg1967, Mermin1966}. However, since discrete lattice symmetries \emph{can} be broken at finite temperatures, Potts universality can still occur via ordering of a composite magnetic order parameter that preserves SO(3) but breaks a lattice symmetry. This leads to the intriguing situation that the appearance of long-range discrete order is driven by fluctuations of an underlying continuous degree of freedom.
Phase transitions involving vestigial order parameters that survive partial melting of a primary order have been widely studied in magnetic, charge density wave and superconducting systems~\cite{Chandra1990, Babaev2004, Fradkin2010, Svistunov2015, Fernandes2019}, offering a natural explanation for the complexity of phase diagrams based on symmetry alone.
Emergent $\mathbb{Z}_3$ Potts order was previously found in fully antiferromagnetic SO(3)-invariant spin models on the honeycomb, kagome and triangular lattice \cite{Mulder2010, Li2022, Little2020, Haley2020}, and it was also reported in coupled clock models~\cite{drouin-touchetteEmergentPottsOrder2022}, cold atomic gases~\cite{Jin2021}, twisted bilayer graphene~\cite{Fernandes2020,caoNematicityCompetingOrders2021} and other unconventional superconductors~\cite{Hecker2018, Cho2020,nieChargedensitywavedrivenElectronicNematicity2022, Hecker2022}. Here, we show that it also emerges in SO(3) symmetric magnets with \emph{mixed} ferro- and antiferromagnetic interactions defined on various stacked lattice designs. Our work thus largely extends the material space for the experimental realizations of this phenomenon.  
\begin{figure}[b]
\centering
\includegraphics[width=\linewidth]{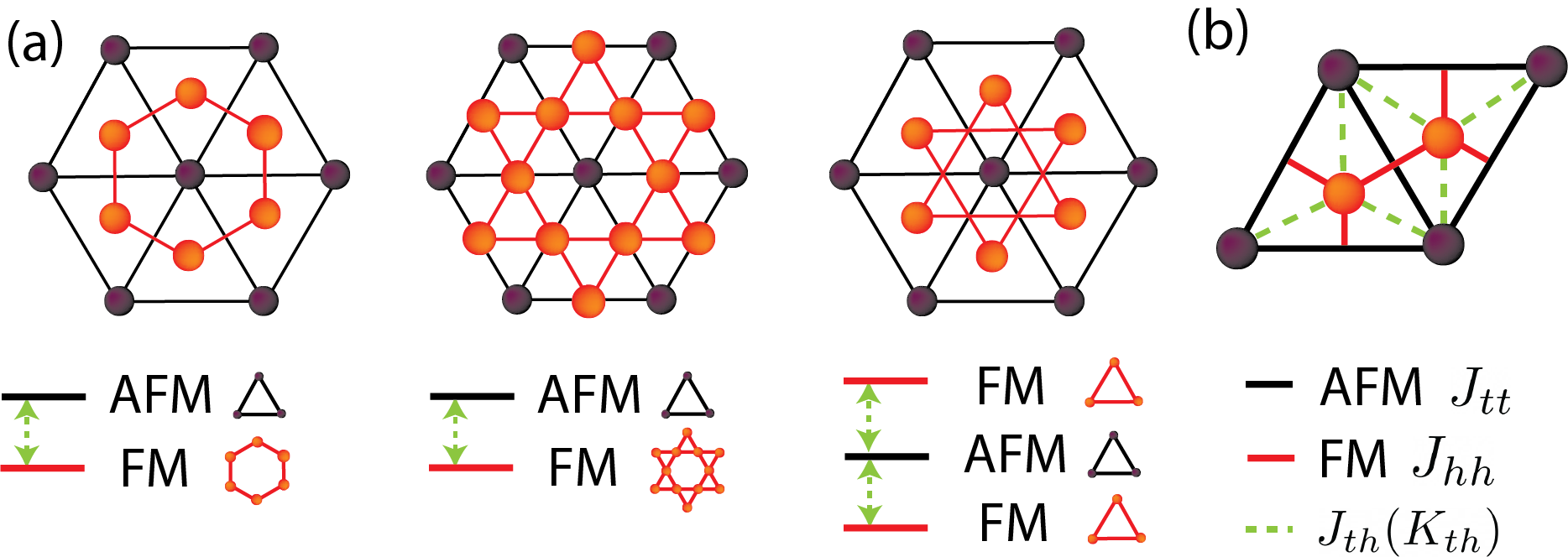}
\caption{(a) Stacked 2D lattice designs of SO(3)-invariant spin models with nearest-neighbor couplings [see Eq.~\eqref{Eq:Hamiltonian}]. They contain an AFM triangular layer (black, $J_{tt} > 0$) coupled to a sixfold symmetric FM layer (red, $J_{hh} < 0$). Different layers are coupled via bilinear $J_{th}$ and biquadratic $K_{th}$ interactions (green). Lattice designs from left to right are triangular-honeycomb, triangular-kagome and ABC stacked triangular. 
(b) Unit cell of the triangular-honeycomb (windmill) lattice with couplings indicated.}
\label{Fig:lattice}
\end{figure}

We design SO(3)-symmetric spin models on stacked 2D lattices that exhibit emergent $\mathbb{Z}_3$ Potts phase transitions at finite temperatures. The models include nearest-neighbor bilinear and biquadratic exchange interactions, where the bilinear interactions take opposite signs on different sublattices. As shown in Fig.~\ref{Fig:lattice}, the models contain a frustrated AFM triangular sublattice coupled to a sixfold symmetric FM layer. As a function of the biquadratic coupling, our models effectively interpolate between AFM Heisenberg and Ising models on a triangular lattice in a magnetic field, yet in a fully SO(3) invariant setup.

Materials that realize 2D $\mathbb{Z}_3$ Potts order in elementary \emph{discrete} degrees of freedom have been the subject of intense studies over the years. Examples are inert gases that are adsorbed on graphite substrates~\cite{Bretz1977, Tejwani1980, Vilches1980, Horn1978, Feng1990, Berker1978, Wiechert1991} and realizations in liquid crystals~\cite{Stoebe1992, Huang1993, Stoebe1994, Jin1995, Ou1997, Lin2000}. 
More recent directions include synthetic matter platforms such as Rydberg atom lattices~\cite{Keesling2019} and cold atoms~\cite{Jin2021}. Another possibility to  experimentally realize three-state Potts models are systems that exhibit \emph{emergent} Potts-nematic order, where the order parameter is a composite object~\cite{Xu2018, Venderbos2018, Fernandes2020, Cho2020, Little2020, Islam-arXiv-2022}. 
Here, we identify a new family of continuous SO(3)-invariant Heisenberg spin models on stacked lattice designs that host three-state Potts-nematic order. 

Recently, a number of stacked magnetic materials realizing the required lattice geometry and containing frustrated triangular layers have been found, including the
triangular-honeycomb lattice material K$_2$Mn$_3$(VO$_4$)$_2$CO$_3$~\cite{Garlea2019}, the triangular-kagome lattice film $\mathrm{Mg}{\mathrm{Cr}}_{2}{\mathrm{O}}_{4}$~\cite{Wen-PRB-2020}, and the stacked triangular lattice materials CaMn$_2$P$_2$~\cite{Islam-arXiv-2022} and Fe$_{1/3}$NbS$_2$ \cite{Little2020, Haley2020} with tunable magnetic phases under intercalation~\cite{Wu2022}. While the stacked lattices in these materials agree with the ones we propose in Fig.~\ref{Fig:lattice}, the exchange interactions in these materials are different, for example, the honeycomb layer in K$_2$Mn$_3$(VO$_4$)$_2$CO$_3$ exhibits AFM rather than FM interactions. We are not aware of a material candidate that exactly realizes the spin model we introduce and study below. Since materials with mixed ferro- and antiferromagnetic interactions are rather common, we hope, however, that our work will stimulate efforts to find material candidates with stacked AFM triangular and FM sixfold symmetric layers. In addition, stacking two-dimensional van der Waals magnets~\cite{Vaz2008, makProbingControllingMagnetic2019, Jenkins2022} poses an alternative route to the realization of the proposed spin models. 

This paper is organized as follows.  In Section~\ref{Sec:Model}, we introduce the model we consider, a rotationally-symmetric bilayer spin Hamiltonian with bilinear and biquadratic exchange interactions, which we investigate for the case of classical spins here. The zero-temperature phase diagram is mapped out in Section~\ref{Sec:Zero-temperature_PD}, showing a large region in which the UUD state is the ground state. Our key results are presented in Section~\ref{Sec:FiniteT_PD}, which includes the finite-temperature phase diagram that hosts a region with Potts-nematic order, as well as a scaling analysis to establish the Potts criticality of the phase transition. It also includes a discussion of domain walls, that are responsible for disordering the $\mathbb{Z}_3$ Potts UUD phase. In Section~\ref{Sec:Fluctuations}, we discuss the role of quantum and thermal fluctuations in promoting biquadratic exchange via an order-by-disorder mechanism in a model with purely bilinear Heisenberg interactions. Finally, we present conclusions in Section~\ref{Sec:Conclusions} and the details of a few calculations in the Appendices. 


\section{Bilayer spin model}
\label{Sec:Model}
We consider SO(3)-symmetric bilayer spin Hamiltonians of the form  
\begin{align}
\mathcal{H}=\sum_{\substack{\langle i,j\rangle_{\alpha\beta}}
}J_{\alpha\beta}\mathbf{S}_{\alpha i}\cdot\mathbf{S}_{\beta j}+\sum_{\substack{\langle i,j\rangle_{th}}
}K_{th}\left(\mathbf{S}_{t i}\cdot\mathbf{S}_{h j}\right)^{2}.
\label{Eq:Hamiltonian}
\end{align}
Here, $\alpha, \beta\in \{t,h\}$ labels the different layers with $t$ referring to the AFM triangular layer ($J_{tt} > 0$) and $h$ to the FM coupled sixfold symmetric layer ($J_{hh} < 0$). We consider both signs of the sublattice couplings $J_{th}$ and $K_{th}$. The summation runs over nearest-neighbor pairs of spins on sublattices $\alpha, \beta$. In the following, we focus on classical spin models for which $\mathbf{S}_{\alpha i}$ are unit-length vectors. While we include a biquadratic interlayer interaction $K_{th}$ in the model, we show later in Sec.~\ref{Sec:Fluctuations} that the effects of such a coupling emerge naturally in a purely bilinear model from quantum and thermal fluctuations via an order-by-disorder mechanism~\cite{shenderAntiferromagneticGarnetsFluctuationally1982,Henley_PRL_1989, henleySemiclassicalEigenstatesFourSublattice1998, jacobsFluctuationinducedPhaseTransverse1998}. This reduces the requirements for the experimental realization of the model.

For concreteness, we study the triangular-honeycomb lattice design (also known as the windmill lattice~\cite{Orth2012, Orth2014}) in the following, which is shown on the left in Fig.~\ref{Fig:lattice}(a). We note that the fully AFM version of the windmill model has previously been studied and found to host an emergent $\mathbb{Z}_6$ order parameter~\cite{Orth2012, Orth2014, Jeevanesan2015}. Our general results also apply to the other lattice geometries, triangular-kagome and ABC-stacked triangular lattices, if one applies a simple rescaling of the FM $J_{hh}$ coupling. The triangular-kagome lattice model [middle panel in Fig.~\ref{Fig:lattice}(a)] is described by Eq.~\eqref{Eq:Hamiltonian} with rescaled coupling $J_{hh}' = \frac{3}{4} J_{hh}$. Here, $J_{hh}'$ denotes the FM coupling in the kagome layer). The ABC stacked triangular lattice [right panel in Fig.~\ref{Fig:lattice}(a)] is described by Eq.~\eqref{Eq:Hamiltonian} with $J_{hh}' = \frac{1}{2} J_{hh}$ with $J_{hh}'$ being the coupling in each FM triangular layer. The other couplings in the triangular-kagome and ABC-triangular models are the same as in the windmill lattice model. 

 We note that here we focus on the minimal classical microscopic model that leads to the emergent Potts physics in the lattices we are studying.
We, therefore, do not include further-range spin-spin interactions or additional intralayer nearest-neighbor biquadratic exchange interactions. A biquadratic coupling on the honeycomb layer, $K_{hh}<0$, would not add anything new to the model as a collinear arrangement of the spins is already preferred by the FM interaction $J_{hh} < 0$. In contrast, an easy-plane biquadratic exchange $K_{hh}>0$ would compete with $J_{hh}$ and lead to a more complex phase diagram if it is sufficiently strong. A biquadratic term on the triangular lattice, $K_{tt}$, would either favor the collinear UUD state for $K_{tt} < 0$ or a non-coplanar umbrella state for $K_{tt} > 0$. Since the parameter $K_{th}$ plays a similar role as we show below, we do not include $K_{tt}$ in our minimal model. 

\section{Zero-temperature classical phase diagram}
\label{Sec:Zero-temperature_PD}
In this Section, we establish the zero-temperature phase diagram of the model in Eq.~\eqref{Eq:Hamiltonian} as a starting point for the finite temperature study. We focus on the regime of FM $J_{hh} <0$ and AFM $J_{tt}>0$ and determine the ground state (GS) phase diagram using two complementary techniques: we first employ a variational ansatz, which assumes a three-sublattice periodicity on the triangular lattice and a uniform state on the honeycomb lattice. In addition to the ground state, we obtain analytical expressions for their energies and the phase boundaries. We then confirm the variational results using low-temperature classical Monte Carlo (MC) simulations. 


\subsection{Variational phase diagram}
\label{subsec:variational_phase_diagram}

\begin{figure*}[t]
\includegraphics[width=\linewidth]{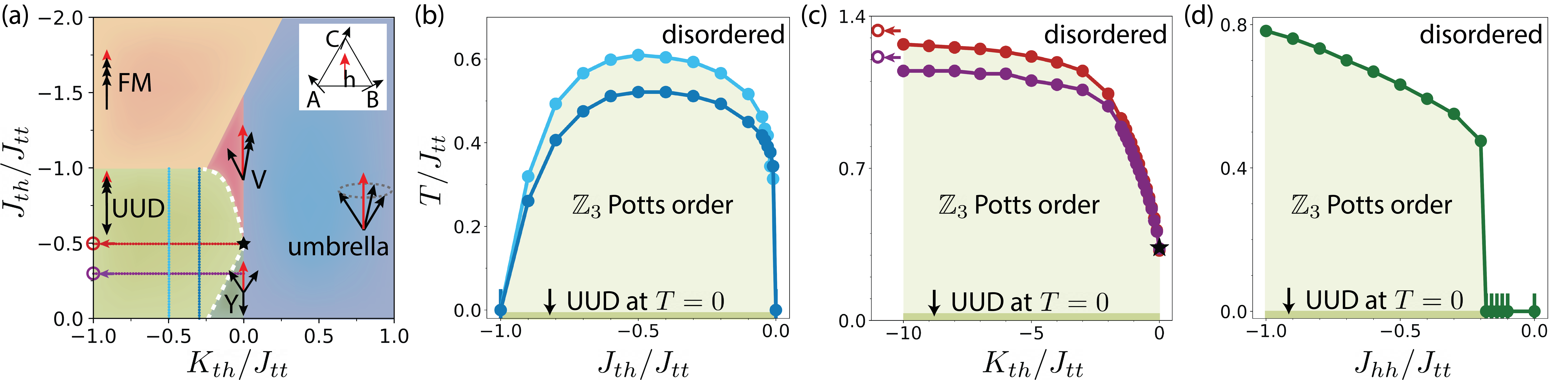}
\caption{(a) Zero-temperature variational phase diagram for FM $J_{hh}$ and AFM $J_{tt}$ in the regime $|J_{hh}| \gtrsim J_{tt}$ as a function of $K_{th}$ and $J_{th}$. The phase diagram is obtained using a variational ansatz with three types of triangular spins (A,B,C) (black) and one type of honeycomb spins ($h$) (red) [see inset]. Arrows illustrate classical ground states. The dashed white line is obtained from classical MC calculations for $J_{hh}=-J_{tt}=-1$. Colored lines correspond to one-dimensional cuts shown at finite temperatures in panels (b-c). 
(b) $\mathbb{Z}_3$ Potts-nematic transition temperature $T_c$ as a function of $J_{th}$ and $T$ above UUD phase at fixed $K_{th}/J_{tt}=-0.3$ (dark blue) and $K_{th}/J_{tt}=-0.5$ (light blue). Other parameters are $J_{tt} = 1$, $J_{hh} = -1$. 
(c) Potts $T_c$ as a function of $K_{th}$ for fixed $J_{th}/J_{tt} = -0.3$ (purple) and $J_{th}/J_{tt}=-0.5$ (red). Other parameters are $J_{tt} = 1$, $J_{hh} = -1$. 
The star shows Potts $T_c$ in the Heisenberg triangular AFM in a magnetic field at the UUD point~\cite{Seabra2011}, and empty circles are Potts $T_c$ for the Ising triangular AFM in corresponding magnetic fields~\cite{kinzelPhenomenologicalScalingApproach1981, qianCriticalFrontierTriangular2004}.
(d) Potts $T_c$ as a function of $J_{hh}$ for fixed $J_{th}=-0.5$,  $K_{th}=-1$ and $J_{tt} = 1$. The values for Potts $T_c$ are obtained using MC simulations from the crossing of the Binder cumulant $U_2$ of $\boldsymbol{m}_3$.
}
\label{Fig:PD}
\end{figure*}

We determine the classical ground state phase diagram for FM $J_{hh} <0$ and AFM $J_{tt}>0$ assuming a variational ansatz that considers one honeycomb (h) and three triangular (A,B,C) sublattices [see inset in Fig.~\ref{Fig:PD}(a)]. This ansatz is well justified in the regime $|J_{hh}| \gg J_{tt} > |J_{th}|, |K_{th}|$. The analytical form of the ansatz is given in Appendix~\ref{SM_var_ansatz}. To obtain the GS phase diagram, we first numerically minimize the variational energy for fixed interaction parameters $J_{th}/J_{tt}$ and $K_{th}/J_{tt}$. The contribution from parameter $J_{hh}$ is constant in the variational manifold. We start the minimization from $20$ different random initial sets of variational angles describing the spin directions and keeping the lowest energy solution. We identify five classes of ground states that appear in the phase diagram: FM, UUD, umbrella, Y, and V state. Considering the symmetries of these phases, we find simpler expressions for the variational energies using fewer angles (see Appendix~\ref{SM_var_ansatz}). By comparing different energies, we also obtain an analytical expression of the phase boundaries. Details of this calculation and the explicit expressions are provided in Appendix~\ref{SM_var_ansatz}.

The resulting ground state phase diagram as a function of nonzero $J_{th}/J_{tt}$ and $K_{th}/J_{tt}$ is shown in Fig.~\ref{Fig:PD}(a). The form of the ansatz assumes that $|J_{hh}| \gtrsim J_{tt}$ and we confirm below using MC simulations that it holds also for $|J_{hh}| = J_{tt}$. The phase diagram contains five different phases with spin arrangements sketched in Fig.~\ref{Fig:PD}(a), where the red spin refers to the direction of the uniform honeycomb layer and the three black spins denote the directions of the triangular spins on the three sublattices A, B, C. The symmetries that are broken in the ordered states are described by the order parameter manifolds of degenerate ground states, which are given in Table~\ref{tab:order_param}. To discuss the phase diagram, we first notice that $K_{th} < 0$ favors coplanar and collinear phases (UUD, V, Y, FM), while $K_{th} > 0$ prefers non-coplanar phases (umbrella). Our focus in the following is on the $K_{th} < 0$ regime, in particular the collinear UUD region, which is the ground state for $0 > J_{th}/J_{tt}> -1 $ and sufficiently negative $K_{th}$. For small $|K_{th}|/J_{tt}$, the system undergoes a sequence of transitions as $|J_{th}|/J_{tt}$ increases from Y to UUD to V and finally to the FM phase. While we focus on FM $J_{th} < 0$ here, we note that the phase diagram for AFM $J_{th}$ is easily obtained by inverting the spins on the honeycomb lattice $\mathbf{S}_h \rightarrow - \mathbf{S}_h$. 
\begin{table}[b]
    \centering
    \begin{tabular}{*{2}{c}} 
     \toprule
     Classical ground state & Order parameter manifold \\
     \midrule
     FM & SO(3) \\ 
     UUD & SO(3) $\times$ $\mathbb{Z}_3$ \\
     Y & SO(3) $\times$ U(1) $\times$ $\mathbb{Z}_3$  \\
     V & SO(3) $\times$ U(1) $\times$ $\mathbb{Z}_3$   \\
     Umbrella & SO(3) $\times$ U(1) $\times$ $\mathbb{Z}_2$ \\
     \bottomrule
    \end{tabular}
    \caption{Order parameter manifolds describing the symmetries that are broken in the different classical ground states. It also corresponds to the ground state degeneracy. 
    Apart from the additional SO(3) symmetry, these are identical to ones found for the triangular Heisenberg AFM in a magnetic field~\cite{Starykh2015}.}
    \label{tab:order_param}
\end{table}

\subsection{MC low-temperature phase diagram}
\label{SM_numerical_PD}

\begin{figure}[t]
    \centering
    \includegraphics[width=0.95\columnwidth]{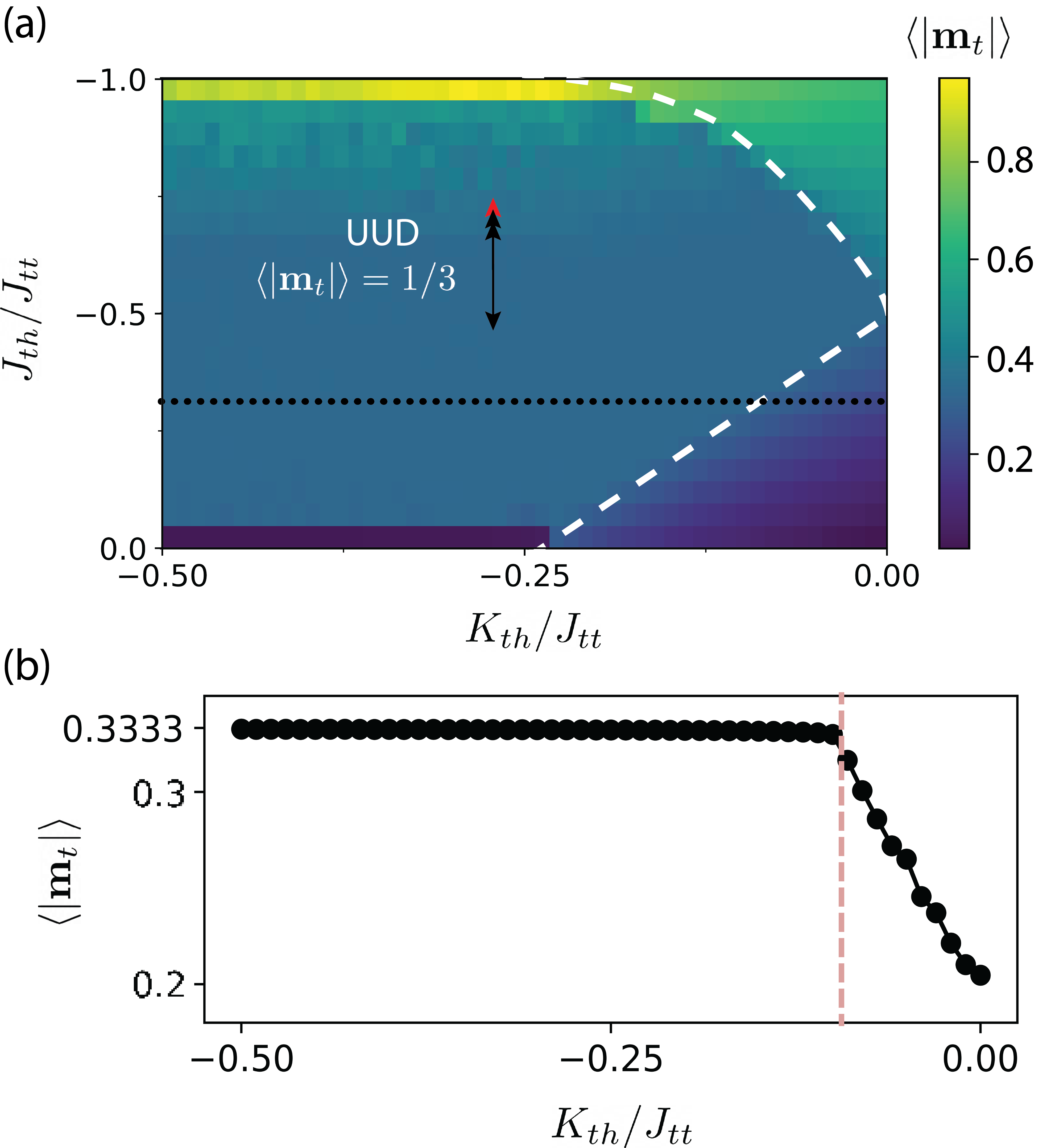}
    \caption{(a) Low-temperature MC phase diagram showing the absolute value of the magnetization on the triangular sublattice as a function of $K_{th}/J_{tt}$ and $J_{th}/J_{tt}$. Different phases can be read off from the value of $|\langle \mathbf{m}_t|\rangle$, as explained in the text. 
    The dashed white lines illustrate the boundaries of the UUD phase with other coplanar phases as obtained by $|\langle |\mathbf{m}_t|\rangle - 1/3| > 10^{-2}$ [see panel (b)] for $J_{th}/J_{tt} \in (0, -0.75)$. For the range $J_{th}/J_{tt} \in (-0.75, -1)$ where thermal fluctuations are larger, the cutoff used is $|\langle |\mathbf{m}_t|\rangle| - 1/3 > 10^{-1}$. This boundary is also shown on top of the variational phase diagram in Fig.~\ref{Fig:PD}(a). The MC phase diagram is obtained for $J_{tt}=1$, $J_{hh}=-1$ and averaged over 15 logarithmically spaced temperature values $T/J_{tt} \in (0.005, 0.05)$. 
    (b) A cut through the data shown in panel (a) at fixed $J_{th}/J_{tt}=-0.3$ [see dotted black line in (a)]. It shows the absolute value of the magnetization on the triangular sublattice as a function of $K_{th}/J_{tt}$ with the vertical dashed pink line illustrating the boundary between the UUD and the V phase.}
    \label{Fig:UUD_boundaries_MC}
\end{figure}
We confirm the variational phase diagram in Fig.~\ref{Fig:PD}(a) using classical Monte Carlo (MC) simulations at $J_{hh} = - J_{tt}$. We determine the phase boundaries between the Y, V and UUD phases as a function of $J_{th}$ and $K_{th} < 0$ and find that the MC phase boundaries precisely agree with the variational ones (see the MC boundaries as white dashed lines in the variational phase diagram Fig.~\ref{Fig:PD}(a)). We do not find evidence for new phases with a larger periodicity. 

The classical Monte Carlo calculations are performed for a series $400$ temperatures, grouped into two logarithmically spaced sets between $0.05 \leq T/J_{tt} \leq 5.0$ and between $0.3 \leq T/J_{tt} \leq 5.0$. Each MC step includes a local Metropolis MC update and a parallel-tempering move between the ensembles at different temperatures, using $5\times 10^5$ MC steps in total. The first half is discarded for thermalization and we perform measurements only in the second half.
To determine the low-temperature phase diagram in the range $-0.5 \leq K_{th}/J_{tt} \leq 0$ and $-1.0 \leq J_{th}/J_{tt}  \leq 0$, we run MC simulations on a detailed coupling parameter grid with spacings $\Delta K_{th}/J_{tt}=0.01$ and $\Delta J_{th}/J_{tt}=0.05$, and use fixed $J_{hh} = -1, J_{tt} = 1$. The calculations are obtained for linear system sizes of $L=24$ unit cells, corresponding to a total number of spins $N = 3 L^2 = 1728$.

To distinguish the different phases, we analyze in Fig.~\ref{Fig:UUD_boundaries_MC} the absolute value of the magnetization on the triangular sublattice
\begin{equation}
\langle |\mathbf{m}_t| \rangle = \left\langle \frac{1}{N_t} \left|\sum_i \mathbf{S}_{ti} \right| \right\rangle \,.
\end{equation}
Here, $i$ runs over the $N_t = L^2$ spins on the triangular sublattice, and $\langle \cdot \rangle$ denotes both the MC average and average over 15 temperatures,  $T/J_{tt} \in (0.005, 0.05)$. Note that while $\langle |\mathbf{m}_t| \rangle$ vanishes in the thermodynamic limit at finite temperatures, it converges to a finite value 
for finite system sizes at low temperatures. We can thus use MC simulations at finite temperature to determine the ground state value of $\langle |\mathbf{m}_t| \rangle$ at $T=0$.
We also compute the chirality of spins on the triangular sublattice 
\begin{equation}
    \chi_c = \left\langle \frac{1}{N_p} \sum_i (\mathbf{S}_{Ai} \times \mathbf{S}_{Bi}) \cdot \mathbf{S}_{Ci} \right\rangle \,,
    \label{eq:spin_chirality}
\end{equation}
where $N_p=2L^2$ is the number of the triangular plaquettes, and confirm that it vanishes, $\chi_c = 0$, for all phases shown in Fig.~\ref{Fig:UUD_boundaries_MC}(a). These phases  at $K_{th} < 0$ are thus either coplanar or collinear. We also confirm that non-coplanar symmetric umbrella states with finite chirality appear for $K_{th}>0$, but these are not the focus of this work.  

Fig.~\ref{Fig:UUD_boundaries_MC} illustrates the separation of the UUD phase for which $\langle|\mathbf{m}_t|\rangle= 1/3$ from other phases for which $\langle|\mathbf{m}_t|\rangle \neq 1/3$. The white dashed lines indicate the phase boundary obtained from cuts at fixed $J_{th}$ such as the one shown in Fig.~\ref{Fig:UUD_boundaries_MC}(b). As noted above, the MC phase boundaries between the UUD and the V and Y states are also included in Fig.~\ref{Fig:PD}(a), validating the variational approach. Note that at $K_{th}=0$, the model can be related to the known case of the triangular Heisenberg AFM in a magnetic field~\cite{Starykh2015}, because the ferromagnetically ordered honeycomb spins (at $T=0$) act as an external magnetic field on $\mathbf{S}_{ti}$. Like the Heisenberg AFM in a magnetic field, our model exhibits the V state for $\langle|\mathbf{m}_t|\rangle > 1/3$, the UUD state for $\langle|\mathbf{m}_t|\rangle = 1/3$ and the Y state for $\langle|\mathbf{m}_t|\rangle < 1/3$, yet in a fully SO(3)-invariant model - of course, SO(3) is broken at $T=0$ by the FM order on the honeycomb sublattice.


\section{Finite-temperature phase diagram and Potts-nematic transition}
\label{Sec:FiniteT_PD}
In this Section, we examine the finite-temperature phase diagram above the UUD ground state, where the system undergoes a $\mathbb{Z}_3$ Potts-nematic transition. We put an emphasis on exploring different parameter limits, where we can make connections and point out differences with the Ising and Heisenberg models in magnetic fields. We also  numerically establish the universality of the phase transition. We start from the UUD phase at $T=0$ and characterize the emergent $\mathbb{Z}_3$ Potts-nematic order that melts at a finite transition temperature $T_c$. 

At finite temperature, only the discrete $\mathbb{Z}_3$ part of the UUD order parameter can exhibit long-range order since the continuous SO(3) degrees of freedom remain disordered. The $\mathbb{Z}_3$ breaking corresponds to a breaking of a discrete lattice symmetry: in this case, it is translational symmetry, leading to a tripling of the unit cell. To identify the $\mathbb{Z}_3$ broken phase, we construct a composite $\mathbb{Z}_3$ Potts-nematic order parameter $\mathbf{m}_3 = m_3 e^{i \theta}$ as
\begin{equation} 
m_3 e^{i \theta} = -\frac{1}{2 N_p} \sum_{i=1}^{N_p} \left(\mathbf{S}_{Ai}+\mathbf{S}_{Bi}e^{\frac{2\pi i}{3}}+\mathbf{S}_{Ci}e^{-\frac{2\pi i}{3}}\right)\cdot \mathbf{S}_{hi} \,.
\label{Eq:order_parameter}
\end{equation}

\begin{figure}[t]
\centering
\includegraphics[width=\linewidth]{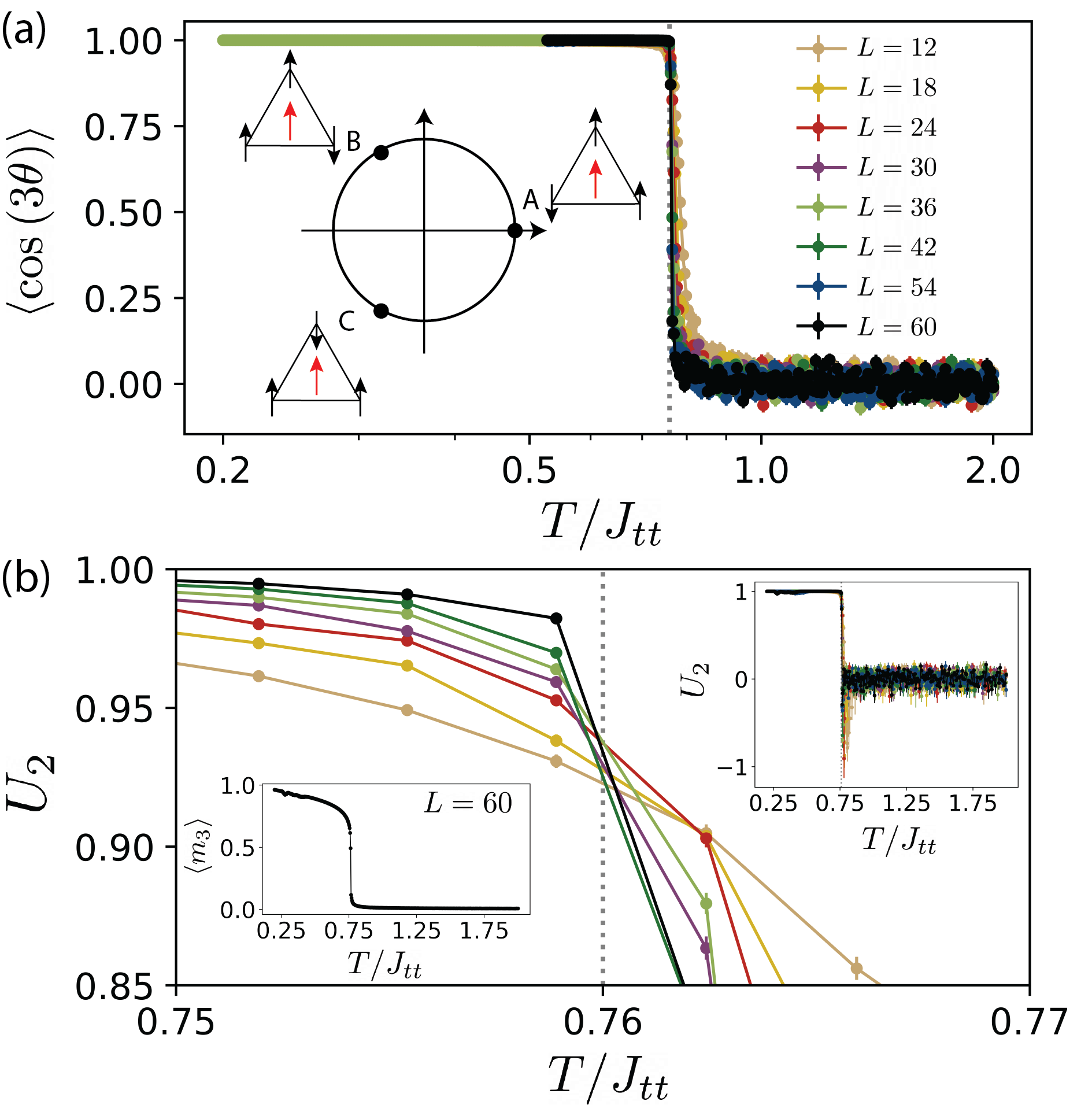}
\caption{
(a) MC results of Potts-nematic order parameter phase $\langle \cos(3\theta)\rangle$ as a function of temperature for different system sizes up to a linear size $L=60$. The spin coupling parameters are $J_{hh} = - J_{tt} = -1$, $J_{th} = -0.3$, $K_{th} = -1$, corresponding to an UUD ground state. The phase shows a sharp continuous phase transition from a disordered to a $\mathbb{Z}_3$ ordered phase, where $\theta = 0, \pm \frac{2 \pi}{3}$ falls into one of three degenerate minima. The grey vertical line indicates $T_c$ determined from the Binder cumulant $U_2$ shown in the panel (b). The inset illustrates the three degenerate configurations of the Potts-nematic order parameter $\boldsymbol{m}_3 = m_3 e^{i\theta}$ in the ordered phase. 
(b) Binder cumulant $U_2$ of $\boldsymbol{m}_3$ as a function of temperature $T$ for different system sizes $L$. The crossing of $U_2$ for different $L$ is used to determine $T_c$ in Fig.~\ref{Fig:PD}. The upper inset shows $U_2$ over a larger range of temperatures. The lower inset shows the sharp onset of a finite magnitude of $\langle m_3 \rangle$ versus $T$ for $L=60$. Coupling parameters are as in (a).}
\label{Fig:order_parameter}
\end{figure}

The summation runs over all $N_p=2L^2$ triangular plaquettes consisting of three triangular and one honeycomb spin [see inset of Fig.~\ref{Fig:PD}(a)]. In the $T=0$ ground state UUD phase, one finds $m_3=1$ and $\theta = 0, \pm \frac{2 \pi}{3}$ corresponding to the three degenerate configurations, which are illustrated in Fig.~\ref{Fig:order_parameter}(a). Unlike the corresponding Potts order parameter of the Heisenberg AFM in an external field~\cite{Seabra2011}, this composite Potts nematic order parameter $\boldsymbol{m}_3$ probes the \emph{relative} alignment of triangular and honeycomb spins. To relate our results to the models in the field, one can interpret the honeycomb spin direction $\mathbf{S}_{hi}$ as a spatially fluctuating coordinate system $S^z$-direction for the triangular spins. While the honeycomb spin correlation length $\xi_h$ remains finite at $T>0$, it becomes exponentially larger than the lattice scale when $T < |J_{hh}|$. The resulting honeycomb magnetization that the triangular spins experience is therefore a slowly varying function of position. 

In the following, we focus on the finite temperature $\mathbb{Z}_3$ Potts-nematic transition above the UUD ground state, which covers a large region of the phase diagram. 
The Y, V, and umbrella phases that appear on the zero-temperature phase diagram, all exhibit an additional $U(1)$ symmetry [see Table~\ref{tab:order_param}], which results in a more complex finite-T behavior in which the Potts transition competes with a KT transition of the $U(1)$ degree of freedom. Elucidating the sequence of the two transitions and performing a scaling analysis above these phases requires a systematic numerical study that goes beyond the scope of this work and is left for a future study.

\subsection{Numerical phase diagram of three-state Potts-nematic transition}
We use MC simulations to determine the finite temperature phase diagram above the UUD ground state. In Fig.~\ref{Fig:PD}(b-d), we present  different two-dimensional cuts through the phase diagram that show the behavior of the transition temperature $T_c$ into the Potts-nematic ordered phase as a function of $J_{th}, K_{th}, J_{hh}$, keeping two of them fixed in each panel.
We obtain $T_c$ from the crossing of the Binder cumulant 
\begin{equation}
U_2 = 2 \Bigl(1 - \frac{\langle |\boldsymbol{m}_3|^4\rangle}{2 \langle |\boldsymbol{m}_3|^2\rangle^2}\Bigr)\,,
    \label{eq:Binder_U2_m3}
\end{equation}
derived in Appendix~\ref{SM_Binder_cumulant} and calculated at different system sizes up to $L = 36$.  

As a function of intersublattice coupling $J_{th}$ and fixed $K_{th} < 0$, the transition temperature $T_c$ exhibits a dome-like behavior, shown in Fig.~\ref{Fig:PD}(b). Tuning $J_{th}$ acts similarly to tuning an external magnetic field in a triangular Ising AFM~\cite{Metcalf1973, kinzelPhenomenologicalScalingApproach1981}, yet in a fully SO(3) preserving way. While $T_c$ approaches zero smoothly as $J_{th} \rightarrow -1$, where the underlying UUD ground state changes into the FM state via a second-order transition, $T_c$ drops abruptly to zero when $J_{th} \rightarrow 0^+$. Numerically, we find that $T_c\to0$ for $|J_{th}|<0.005$.
In Sec.~\ref{SM_Peierls_argument}, we show that this vanishing of the critical temperature is driven by the appearance of domain walls whose energy approaches zero in the given limits.
While the behavior as $J_{th} \rightarrow 0^+$ may be indicative of a first-order phase transition, the related triangular Ising AFM in the field is known to undergo a KT transition at small temperature and field with similar features~\cite{qianCriticalFrontierTriangular2004}.

\begin{figure*}[t]
\includegraphics[width=\linewidth]{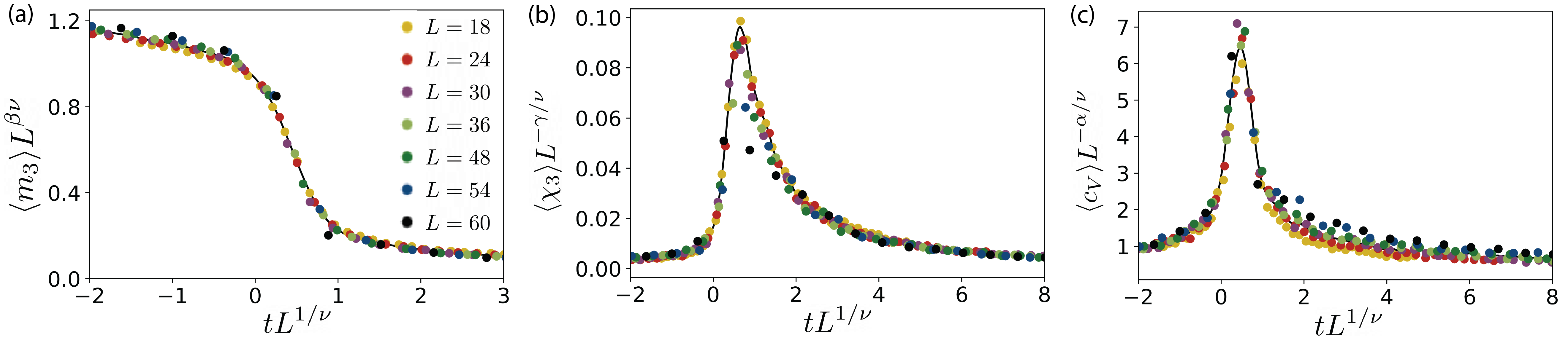}
\caption{Finite size scaling analysis for (a) $\mathbb{Z}_3$ Potts magnetization $\langle m_3 \rangle$ (b) $\mathbb{Z}_3$ Potts susceptibility $\langle \chi_3 \rangle$ and (c) the specific heat $\langle c_V \rangle$ for linear system sizes up to $L=60$ unit cells using critical exponents from $\mathbb{Z}_3$ Potts universality class, where $t=(T-T_c)/T_c$. Black lines denote scaling functions extracted from a fit of the collapsed data. The MC calculations are performed at finite temperatures above the UUD ground state for $J_{tt}=1$, $J_{hh}=-1$, $J_{th}=-0.3$, and $K_{th}=-1$.}
\label{Fig:scaling_collapse}
\end{figure*}

The relation of our model to the triangular Ising and Heisenberg AFMs can be further studied by tuning the biquadratic coupling $K_{th}$, which tunes the ``Isingness" of the spins 
as shown in Fig.~\ref{Fig:PD}(c). This is the case as large $|K_{th}|$ favors the relative collinear orientation between the triangular and honeycomb spins. The mapping of the model, Eq.~\eqref{Eq:Hamiltonian} to the Ising model on the triangular lattice in a magnetic field for large $K_{th}$ is derived in Appendix~\ref{Appendix:Mapping_Ising}. In the Heisenberg limit, $K_{th} \rightarrow 0^-$, $T_c$ approaches the known value of the Potts transition in the Heisenberg AFM~\cite{Seabra2011}, indicated by the star for the case of $J_{th}/J_{tt} = -0.5$. Potts $T_c$ increases when $K_{th}$ becomes more negative and the system effectively behaves more Ising-like, while still being fully SO(3) invariant. For large negative $K_{th}$, $T_c$ saturates at a value that approaches the known transition temperature of the triangular Ising AFM in an external magnetic field~\cite{Metcalf1973,qianCriticalFrontierTriangular2004}. 

Being fully SO(3) invariant, our model allows for a new tuning parameter that is absent in the triangular AFMs in external fields: \textit{the rigidity} of the magnetic field that the triangular spins experience given by the lengthscale of spin fluctuations on the honeycomb sublattice. This lengthscale is given by the honeycomb spin correlation length $\xi_h$, which is controlled by the ratio $J_{hh}/T$. As shown in Fig.~\ref{Fig:PD}(d), $T_c$ decreases as a function of $J_{hh}$, or decreasing correlation length, and sharply drops to zero at $J_{hh}/J_{tt} = -0.2$. Whether the universality class of this transition is first order or of KT type, like in the case of tuning the field strength, is an open question that deserves further study. We note that the drop in $T_c$ coincides with the position of a GS phase transition out of the UUD state. 

The MC simulation parameters that were used to obtain the phase diagram are as follows. We simulate systems up to a system size of $L=36$ containing a total of $N = 3 L^2=3888$ spins, using periodic boundary conditions (PBC). The simulations employ $10^6$ MC steps, half of which are discarded to let the system thermalize. As previously described, each MC step consists of a Metropolis and a parallel-tempering update. We simulate $400$ temperatures in parallel, grouped into two logarithmically spaced sets between $0.05 \leq T/J_{tt} \leq 5.0$ and between $0.3 \leq T/J_{tt} \leq 5.0$. We choose two sets with different minimal temperatures to get a closely spaced coverage of the phase diagram around intermediate temperatures, where we expect the phase transitions to occur, as well as at low temperatures, where the system approaches the ground state.

\subsection{The universality of the Potts-nematic  transition}
To establish the $\mathbb{Z}_3$ Potts universality class of the nematic phase transition at $T_c$, we perform detailed MC simulations at the parameter set $J_{hh}=-J_{tt}=-1$, $J_{th}=-0.3$, and $K_{th}=-1$. At $T=0$, this parameter point lies deep in the UUD ground state phase. We now consider linear system sizes up to $L=60$ with PBC and use $2 \times 10^6$ MC steps, the first half of which are again discarded for thermalization. We simulate $500$ temperatures in the range $0.02 \leq T/J_{tt} \leq 2.0$ that are spaced logarithmically. 

In Fig.~\ref{Fig:order_parameter}(a), we show the order parameter phase $\langle \cos(3\theta) \rangle$ as a function of temperature for different system sizes $12 \leq L \leq 60$, where $\langle \, \cdot \, \rangle$ denotes the MC average. We observe a sharp transition from zero to one at a temperature that agrees with $T_c$ extracted from the Binder cumulant in panel (b). The transition demonstrates the ordering of the Potts-nematic phase $\theta = 0, \pm \frac{2 \pi}{3}$ into one of the three minima, illustrated in panel (a). In the lower inset of Fig.~\ref{Fig:order_parameter}(b), we plot the magnitude of the Potts-nematic order parameter $\langle m_3 \rangle$ versus temperature at $L=60$, which also undergoes a sharp transition, qualitatively consistent with the 2D $\mathbb{Z}_3$ Potts exponent $\beta = 1/9$. The plot showing the behavior of $\langle m_3 \rangle$ as a function of linear system size $L$ at temperatures around $T_c$ is given in Appendix~\ref{SM_order_param}. In the main part of the panel (b) and the other inset, we show the Binder cumulant for different $L$ from which we extract a transition temperature $T_c = 0.76 J_{tt}$.

In Fig.~\ref{Fig:scaling_collapse} we show a finite size scaling collapse of various Potts nematic observables using the critical exponents from the $\mathbb{Z}_3$ Potts universality class $\alpha = 1/3$, $\beta=1/9$, $\gamma = 13/9$ and $\nu = 5/6$~\cite{Wu1982}. Different panels show the scaling plots of the Potts magnetization $\langle m_3 \rangle$, Potts susceptibility $\langle \chi_3 \rangle = \frac{N}{T} \bigl(\langle m_3^2\rangle - \langle m_3 \rangle^2 \bigr)$ and the specific heat $\langle c_V \rangle = \frac{1}{N T^2} \bigl( \langle \mathcal{H}^2 \rangle - \langle \mathcal{H} \rangle^2 \bigr)$ as a function of the reduced temperature $t=\frac{T-T_c}{T_c}$, where $N$ is the total number of spins. 
We obtained excellent data collapse using only the leading scaling laws with exponents $\beta, \gamma$, and $\alpha$, respectively, without the need to include any subleading terms in the fit. We find the scaling functions (black lines) using a least square weighted regularization, making no assumptions about the scaling functions. This demonstrates that the nematic phase transition at $T_c$ lies in the universality class of the 2D $\mathbb{Z}_3$ Potts transition. The comparison of the obtained and theoretical critical exponents for different system sizes without finite-size correction is shown in Appendix~\ref{SM_scaling_analysis}.

\subsection{Estimation of critical temperature using Peierls' domain wall argument}
\label{SM_Peierls_argument}
\begin{figure*}[t]
    \includegraphics[width=2\columnwidth]{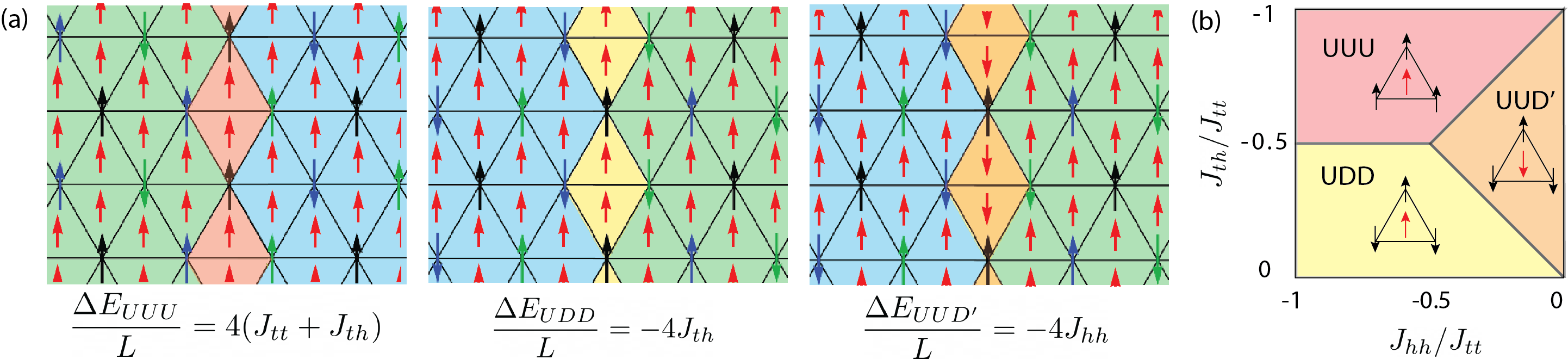}
    \caption{(a) Types of the low-energy short-range domain walls between two different ground states of UUD phase and their corresponding energy per length on interpenetrating triangular (black, green, and blue spins) and honeycomb sublattice (red spins).
    The domain walls consist of plaquettes UUU (red), UDD (yellow), and UUD when the honeycomb spin is flipped (orange). (b) The energetically most favorable types of domain walls as a function of $J_{th}/J_{tt}$ and $J_{hh}/J_{tt}$.}
    \label{Fig:domain_walls}
\end{figure*}

In this Section, we analytically estimate the Potts  transition temperature $T_c$ using Peierls' argument of comparing energy and entropy of introducing domain walls in the ordered state \cite{Peierls1936, Griffiths1964}. Specifically, this argument compares the entropic gain of creating a domain wall between regions with different discrete values of the Potts order parameter with the energy cost associated with the domain wall. Since the underlying spins are continuous, different spin configurations inside of a Potts domain wall are possible and energetically preferred in different parameter regions. First, we identify the energetically most favorable domain walls as a function of $J_{th}/J_{tt}$ and $J_{hh}/J_{tt}$. We focus on short-range \emph{collinear} domain walls between the three different Potts domains, as shown in Fig.~\ref{Fig:domain_walls}(a). This is fully justified for large and negative $|K_{th}|\geq J_{tt}, |J_{th}|$ and $J_{th}<0$. For smaller values of $|K_{th}|$, the energetically most favorable domain walls may have a finite width by balancing gradient energy costs with potential energy. 
In the limit of large negative $K_{th}$, the triangular spins must either be parallel or anti-parallel to the neighboring honeycomb spin. Collinearity then effectively becomes a hard-core constraint and the energy gain of creating a domain wall becomes independent of $K_{th}$.

\begin{table}[b]
\centering
\begin{tabular}{*{3}{c}} 
 \toprule
  DW type & $T_c/J_{tt}$ & Parameters limits where $T_c \to 0$ \\ 
\midrule
 UUU & $2.4(1 + J_{th}/J_{tt})$ & $J_{th}/J_{tt} \to -1^+$, $|J_{hh}|>0.5$ \\ 
 UDD & $2.4 |J_{th}|/J_{tt}$  & $J_{th}/J_{tt} \to 0^-$, $|J_{hh}|>0.5$ \\
 UUD$'$ & $2.4 |J_{hh}|/J_{tt}$ & $J_{hh}/J_{tt} \to 0^-$ \\
 \bottomrule
\end{tabular}
 \caption{Approximate upper bound on the Potts transition temperature $T_c$ and expected parameter limits, where the respective domain wall (DW) energy cost and thus $T_c$ vanish.}
 \label{Tab:Tc_Peierls}
\end{table}

In Fig.~\ref{Fig:domain_walls}(a), we show the different types of short-range collinear domain walls between the different Potts domains and their corresponding energies per length. The short-range collinear domain walls between different Potts domains [or UUD ground states, see Fig.~\ref{Fig:order_parameter}(a)] can be built from UUU or UDD triangular plaquettes, or from a UUD plaquette where the central honeycomb spin is flipped, which we denote as UUD$'$ [see Fig.~\ref{Fig:domain_walls}(b)]. Possible are also domain walls that contain any combination of the three elementary plaquettes, UUU, UDD, UUD'. From the energies of possible short-range domain walls as a function of the interaction parameters, in Fig.~\ref{Fig:domain_walls}(b), we show the energetically most favorable domain wall type as a function of $J_{th}/J_{tt}$ and $J_{hh}/J_{tt}$.

By calculating the energy costs of a domain wall and by estimating the entropy gain $\Delta S$, we can estimate the transition temperature using Peierls' argument as 
\begin{equation}
    T_{c}\sim\Delta E/\Delta S\,.
\end{equation}
The logic is that the order melts when the free energy cost of a domain wall $\Delta F = \Delta E - T \Delta S$ becomes negative. 
As usual, we estimate $\Delta S$ from the number of placements of the domain wall. For domain walls of length $L_{dw}$, the upper bound on the number of the possible placement is $(z-1)^{L_{dw}}$, where $z=6$ is the coordination number of the lattice. Thus, the estimation for the change of entropy in the presence of the domain wall is $\Delta S \lesssim k_B L_{dw} \log{(z-1)} \sim 1.6 \; k_BL_{dw}$. 

Combining the energy cost and the entropy gain of introducing a domain wall, we obtain an approximate upper bound on $T_c$. In particular, when a domain wall energy cost $\Delta E$ approaches zero somewhere in parameter space, we expect that $T_{c} \rightarrow 0$ as well. The parameter limits in which this occurs are shown in Table~\ref{Tab:Tc_Peierls}, and the vanishing of $T_c$ is in agreement with the finite-temperature phase diagram shown in Fig.~\ref{Fig:PD}(b-d).


\section{Biquadratic exchange from fluctuations}
\label{Sec:Fluctuations}
We now show that a biquadratic exchange $K_{th}$ included in the model in Eq.~\eqref{Eq:Hamiltonian} and that is crucial for the development of $\mathbb{Z}_3$ Pott-nematic order is effectively generated by quantum and thermal fluctuations in a purely bilinear model via an order-by-disorder mechanism. To determine the effect of quantum fluctuations, we rely on a known result about the triangular Heisenberg AFM in a magnetic field for which quantum fluctuations have been shown to stabilize the UUD phase~\cite{Chubukov1991}. We can use this result to estimate the biquadratic coupling induced by quantum fluctuations using a mapping between the Heisenberg AFM in a field and our model. 
To determine the effect of thermal fluctuations, we perform a standard calculation that takes into account finite-temperature corrections to the free energy from Gaussian fluctuations. 

\begin{figure*}
    \includegraphics[width=\linewidth]{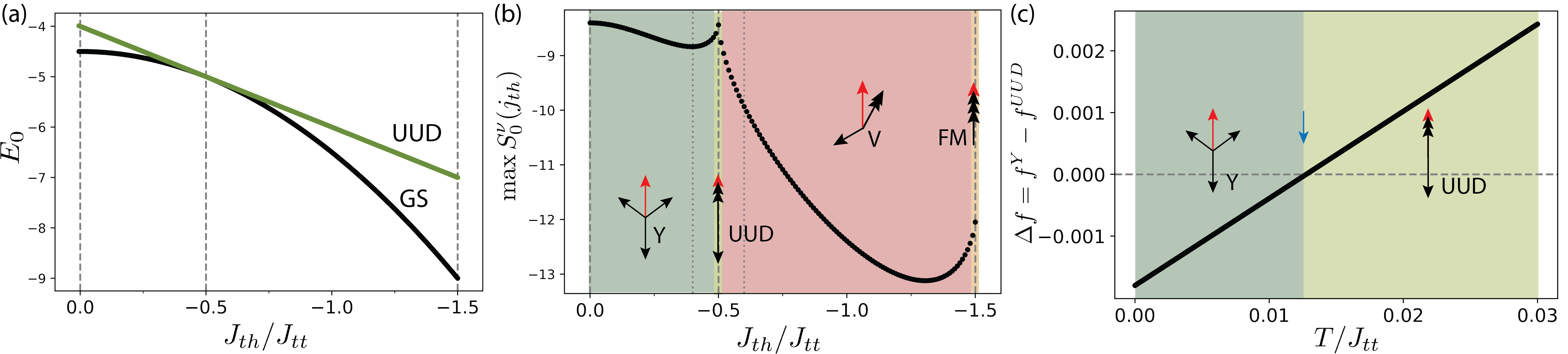}
    \caption{(a) Classical GS energy of model~\eqref{Eq:Hamiltonian} (black) as a function of $J_{th}/J_{tt}$ for fixed $K_{th} = 0$, $J_{hh} = -1$, and $J_{tt} = 1$. The green line shows the energy of the UUD state, which is only part of the highly degenerate GS manifold at $J_{th}/J_{tt} = -1/2$. 
    (b) Maximum value of temperature-independent entropy contribution $S_0^{\nu}(J_{th}/J_{tt}) = - \sideset{}{'}\sum_{\boldsymbol{q}}\ln(\det\hat{A}_{\mathbf{q}}^{\nu})$ over all degenerate GSs as a function of $J_{th}/J_{tt}$ (black). Note that the UUD state is only part of the degenerate GS manifold at $J_{th}/J_{tt} = -1/2$. The states selected by this order-by-disorder mechanism among the degenerate GSs are the coplanar Y state for $|J_{th}/J_{tt}|<0.5$, the collinear UUD state at $|J_{th}/J_{tt}|=0.5$, and the coplanar V state for $|J_{th}/J_{tt}|\in(0.5,1.5)$. In the FM regime for $|J_{th}/J_{tt}|>1.5$ the ground state is unique. The entropy $S_0$ exhibits a peak at $J_{th}/J_{tt}=-1/2$ showing that the entropic contribution is maximal for the UUD state compared to coplanar GSs states nearby.
    (c) Free energy difference $\Delta f = F^{\text{Y}}(j_{th},T) - F^{\text{UUD}}(j_{th}^{(0)}, T)$ as a function of temperature $T$ for fixed $j_{th} = j_{th}^{(0)} + 0.03$ and $j_{th}^{(0)} = -1/2$. This shows that the Y state, which belongs to the degenerate GS manifold, has lower free energy only at low $T$. For $T > 0.012 J_{tt}$ the UUD state exhibits lower free energy due to its larger entropic contribution $S_0$ [see panel (b)]. The UUD state is thus stabilized at finite temperatures via order-by-disorder. }
    \label{Fig:temp_fluct}
\end{figure*}

\subsection{Biquadratic coupling induced by quantum fluctuations}
\label{SM_quantum_fluc}
In this Section, we estimate the size of an effective biquadratic coupling $K^*_{th}$ induced by quantum fluctuations in a purely bilinear model. 
We thus set $K_{th} = 0$ in Eq.~\eqref{Eq:Hamiltonian} and focus on the regime of large FM $|J_{hh}| \gg J_{tt}, |J_{th}|$. Using known results about the quantum triangular Heisenberg AFM in an external magnetic field~\cite{Chubukov1991}, we demonstrate that the quantum fluctuations (via an order-by-disorder mechanism~\cite{shenderAntiferromagneticGarnetsFluctuationally1982,Henley_PRL_1989,henleySemiclassicalEigenstatesFourSublattice1998,jacobsFluctuationinducedPhaseTransverse1998}) induce the same behavior as produced by an effective biquadratic coupling of negative sign: $K^*_{th}<0$, which favors collinear and coplanar states.

When $K_{th}=0$ and at $T=0$, the classical model Eq.~\eqref{Eq:Hamiltonian} can be mapped to the triangular Heisenberg AFM in a magnetic field
\begin{align}
\mathcal{H} = J_{tt} \sum_{\langle i,j \rangle} \mathbf{S}_{ti} \cdot \mathbf{S}_{tj} - hS_t \sum_i S_{ti}^z\,,
\end{align}
where $hS_t = -6 J_{th} S_{h}$ and we have used that the honeycomb spins are all parallel for sufficiently large FM $|J_{hh}| \gg J_{tt}, |J_{th}|$. In the following, we will assume without loss of generality that both spins have the same length $S = S_t = S_h$.
For classical spins, this model exhibits an UUD ground state only for the specific magnetic field value $h^*=3J_{tt}$. It was shown that quantum corrections to order $1/S$, where $S$ is the spin length and $S\rightarrow \infty$ corresponds to the classical limit, stabilize the UUD phase over a finite range of magnetic fields $h \approx h^* \pm \Delta h$ via an order-by-disorder mechanism~\cite{Chubukov1991}:
\begin{align}
\Delta h\approx \frac{1.8}{2S} J_{tt} \,.
\end{align}
From the mapping between the two models, we can relate the magnetic field $h$ to the size of the coupling $J_{th}$ to the honeycomb spins, which exhibit long-range order in the ground state: 
\begin{equation}
    \Delta hS=-6\Delta J_{th} S\,.
\end{equation}
This yields an effective range of couplings $J_{th}$ in which we expect the UUD phase to be stabilized in our model by the quantum fluctuations of triangular spins,
\begin{align}
\Delta J_{th}\approx-\frac{0.3}{2S}J_{tt} \,.
\label{Eq:DeltaJth_Chubukov}
\end{align}
In this estimation we have neglected the fluctuations of the honeycomb spins, which is justified in the regime $|J_{hh}| \gg J_{tt}, |J_{th}|$.

We can now estimate an effective value $K_{th}^{*}$ due to quantum fluctuations that leads to the same stabilization of the UUD phase by comparing to the classical ground state phase diagram of Eq.~\eqref{Eq:Hamiltonian}, shown in Fig.~\ref{Fig:PD}(a). From the analytically known phase boundaries of the UUD phase within the three-sublattice ansatz [see Eqs.~\eqref{boundary1} and~\eqref{boundary2}], we estimate the window of the opening of the UUD phase as a function of $K_{th}$ to be
\begin{align}
\Delta J_{th} \approx 8 K_{th}.
\end{align}
By comparing with Eq.~\eqref{Eq:DeltaJth_Chubukov}, we get the effective biquadratic coupling value
\begin{equation}
    K_{th}^{*} \approx -0.0375 \, \frac{J_{tt}}{2S} \,.
\end{equation}

\subsection{Biquadratic coupling induced by thermal fluctuations}
\label{SM_thermal_fluc}
In this Section, we demonstrate that thermal fluctuations effectively induce a biquadratic coupling $K^*_{th}< 0$ as well. 
Our derivation considers Gaussian thermal fluctuations and we focus around the point $j_{th}^{\left(0\right)}=J_{th}/J_{tt}=-1/2$ for large FM $|J_{hh}| > J_{tt}, J_{th}$
and set $K_{th}=0$ in the classical model~\eqref{Eq:Hamiltonian}. For $J_{hh}/J_{tt}\to-\infty$, the three-sublattice ansatz is exact and
the only constraint on a given triangle for $j_{th} \in\left(-\frac{3}{2},\frac{3}{2}\right)$ is that~\cite{Starykh2015}
\begin{align}
\mathbf{S}_{A}+\mathbf{S}_{B}+\mathbf{S}_{C}&=-2 j_{th} \mathbf{S}_{h} \,.
\label{eq:GS_condition}
\end{align}
At $j_{th}=-1/2$, the UUD state belongs to the massively degenerate classical ground state (GS) manifold that has a magnetization on the triangular lattice equal to $\boldsymbol{m}_t=\mathbf{S}_h/3$, where $\mathbf{S}_h$ describes the FM ordered honeycomb spins. This includes many coplanar states, symmetric umbrella states and a myriad of non-coplanar umbrella-like states. Away from this point at $j_{th} \neq -1/2$, however, the UUD state is no longer part of the ground state manifold, as shown in Fig.~\ref{Fig:temp_fluct}(a). Instead, for $j_{th} < -1/2$ the coplanar V state is part of the (still massively degenerate) manifold and for $j_{th} > -1/2$ the coplanar Y state is part of the degenerate GS manifold. 

It is well known that order-by-disorder favors collinear and coplanar states over non-coplanar ones~\cite{Henley_PRL_1989}. We thus anticipate that among the degenerate ground states, the three states (UUD, V, and Y) will be favored at a finite temperature around $j_{th}^{\left(0\right)}$. In the following, we confirm this expectation by (approximately) calculating the free energy $F = E - TS$ of all collinear and coplanar states in the degenerate ground state manifold, which yields the low-temperature phase diagram. We have also considered some symmetric non-coplanar states, which we always find to have higher free energy, as expected.

To estimate the free energy of the UUD state at finite temperature, we notice that UUD only belongs to the GS manifold for $j_{th} = j_{th}^{\left(0\right)} = -1/2$. At this point, we can calculate the entropic contribution to the free energy by considering Gaussian fluctuations around the UUD state (details shown below), which yields
\begin{equation}
F^{\text{UUD}}\Bigl(j_{th}^{(0)},T\Bigr)= E^{\text{UUD}}\Bigl(j_{th}^{(0)}\Bigr)-T S^{\text{UUD}}\Bigl(j_{th}^{(0)},T\Bigr)\,.
\label{eq:UUD_free_en_jth0}
\end{equation}
Away from this point, $j_{th} \neq j_{th}^{(0)}$, we approximate its free energy  by 
\begin{equation}
F^{\text{UUD}}\left(j_{th},T\right)\simeq E^{\text{UUD}}\left(j_{th}\right)-T S^{\text{UUD}}\left(j_{th}^{\left(0\right)},T\right),
\label{eq:UUD_free_en}
\end{equation}
with $E^{\text{UUD}}\left(j_{th}\right)$ being the energy of the UUD state at $j_{th}$, which is larger than the GS energy at that point. As we will show below, however, the gain in entropy $S^{\text{UUD}}$ can compensate for the higher GS energy such that the free energy of the UUD state is still the lowest one. 
According to our approximation in  Eq.~\eqref{eq:UUD_free_en}, the entropy $S^{\text{UUD}}$ is calculated at $j_{th}^{(0)}=-1/2$ since the UUD state only belongs to the GS manifold at $j_{th}^{(0)}$ and the entropy is calculated from harmonic fluctuations around the ground state. 

At a given $j_{th}$, we compare the free energy of the UUD state as given in Eq.~\eqref{eq:UUD_free_en} to the free energy of states in the GS manifold at that value of $j_{th}$. We label these states by the index $\mu$ in the following, which includes both coplanar and non-coplanar states. Their free energy is given by
\begin{equation}
F^{\mu}\left(j_{th},T\right)\simeq E^{\mu}\left(j_{th}\right)-TS^{\mu}\left(j_{th},T\right)\,,
\label{eq:f0_mu}
\end{equation}
with ground state energy $E^{\mu}\left(j_{th}\right)$ and entropy
$S^{\mu}\left(j_{th}\right)$ arising from Gaussian thermal fluctuations on top
of the ordered state. Since it is generally known that the order-by-disorder mechanism favors collinear and coplanar states over non-coplanar ones, we restrict our analysis to collinear and coplanar states in the GS manifold. We have also explicitly checked that certain types of symmetric umbrella non-coplanar states are never favored and exhibit higher free energy.

When expanding around a given GS, we parameterize the fluctuations in terms of the angles $\delta\theta_{a,i}$
and $\delta x_{a,i}=\sin\theta_{a,i}\,\delta\phi_{a,i}$ for each spin $\boldsymbol{S}_{ai}$ of type $a \in \{A, B, C, \alpha, \beta\}$ at site $i$. This corresponds to spatial fluctuations around a three-sublattice ansatz, where spins on the triangular lattice are labelled as $A,B,C$ and $\alpha, \beta$ refers to the two honeycomb spins in the three-sublattice ansatz. In momentum space this becomes $\psi_{\mathbf{q}}^{T}=\left(\delta\theta_{a, \mathbf{q}},\delta x_{a,\mathbf{q}}\right)$. The Fourier components are summed over half the Brillouin zone, as the fluctuations are real. One can capture Gaussian fluctuations on top of the GS manifold in terms of an effective Hamiltonian
\begin{equation}
\mathcal{H}^{\mu}\left(j_{th}\right)=E^{\mu}\left(j_{th}\right)+\frac{1}{2}\sideset{}{'}\sum_{\mathbf{q}}\psi_{\mathbf{q}}^{\dagger}\hat{A}_{\mathbf{q}}^{\mu}\left(j_{th}\right)\psi_{\mathbf{q}}\,.
\label{eq:Ham_exp}
\end{equation}
The form of the matrix $\hat{A}_{\mathbf{q}}^{\mu}\left(j_{th}\right)$ depends on the GS spin configuration and is analyzed in Appendix~\ref{SM_thermal_fluc_expressions}.
For the UUD state, the effective Hamiltonian is instead written as
\begin{equation}
\mathcal{H}^{\text{UUD}}\left(j_{th}\right)=E^{\text{UUD}}\left(j_{th}\right)+\frac{1}{2}\sideset{}{'}\sum_{\mathbf{q}}\psi_{\mathbf{q}}^{\dagger}\hat{A}_{\mathbf{q}}^{\text{UUD}}\left(j_{th}^{\left(0\right)}\right)\psi_{\mathbf{q}}\,,
\end{equation}
which includes the matrix $\hat{A}_{\mathbf{q}}^{\text{UUD}}\left(j_{th}^{\left(0\right)}\right)$ obtained at $j_{th} = -1/2$. A more accurate approximation would involve computing the fluctuations expanded at $j_{th}$ instead of at fixed $j_{th} = j_{th}^{(0)} = -1/2$. Since the $UUD$ state is not within the GS manifold for $j_{th} \neq -1/2$, however, terms that are linear in $\psi_{\mathbf{q}}$ would be present in the expansion.

We obtain the partition function $Z^{\nu}$ for an effective Hamiltonian $\mathcal{H}^\nu$ by integrating over the Gaussian fluctuations,
\begin{align}
Z^{\nu} =  \frac{e^{-\frac{E_0}{T}}}{(2\pi)^2} \sideset{}{'}\prod_{\mathbf{q}} \int d\psi^{\dagger}_{\mathbf{q}} d\psi_{\mathbf{q}} \;\text{exp} \left[{-\frac{\frac{1}{2} \psi_{\mathbf{q}}^{\dagger} \hat{A}^{\nu}_{\mathbf{q}} \psi_{\mathbf{q}}}{T}}\right]. 
\end{align}
Note that this expression holds for both the UUD state and any of the degenerate GSs $\mu$ at $j_{th}$. 
The integral is straightforward to compute and the free energy can be easily obtained from $F^{\nu} = -T \ln{Z^{\nu}}$ as
\begin{align}
F^{\nu} = E_0^{\nu} \sideset{+T}{'}\sum_{\mathbf{q}}\ln\left(\det\hat{A}_{\mathbf{q}}^{\nu}\right) + N T \ln{\left(2 \sqrt{\frac{\pi}{T}}\right)}\,.
\label{eq:appendix_free_energy}
\end{align}
We obtain the entropy by comparing with Eq.~\eqref{eq:f0_mu} as 
\begin{equation}
    S^{\nu}\left(j_{th}, T\right)=- \sideset{}{'}\sum_{\boldsymbol{q}}\ln\left(\det\hat{A}_{\mathbf{q}}^{\nu}\right) - N \ln\left(2\sqrt{\frac{\pi}{T}}\right) \,.
\end{equation}
In Fig.~\ref{Fig:temp_fluct}(b) we plot the maximum entropy $\max_{\mu}{S^{\mu}\left(j_{th},T\right)}$ considering all states $\mu$ of the GS manifold, dropping the term $-N \ln\left(2\sqrt{\frac{\pi}{T}}\right)$, which is common to all states. Generally, we find that the entropy term favors collinear and coplanar states over non-coplanar ones, with collinear states being most favored. Close to $J_{th}/J_{tt} = -1/2$, $\max_{\mu}{S^{\mu}\left(j_{th},T\right)}$ exhibits a peak at $J_{th}/J_{tt} = -1/2$. This shows that the UUD state has a higher entropic conribution than the nearby coplanar ground states at $J_{th}/J_{tt} \neq -1/2$. This leads to the fact that the UUD state will be favored at finite temperature in a window around $J_{th}/J_{tt} = -1/2$, where the ground state energy difference between the UUD and the other states is small. 

We note that a potential caveat in performing Gaussian expansions around the classical ground state to capture finite temperature fluctuations is the presence of Goldstone modes that are associated with continuous symmetry breaking at zero temperature. However, it is known that while the spin correlation function exhibits an infrared divergence at low temperatures, the free energy is well defined in the low-temperature limit~\cite{Chandra1990, Kawamura1984} and our calculation thus justified. 

\begin{figure}[t]
\centering \includegraphics[width=\linewidth]{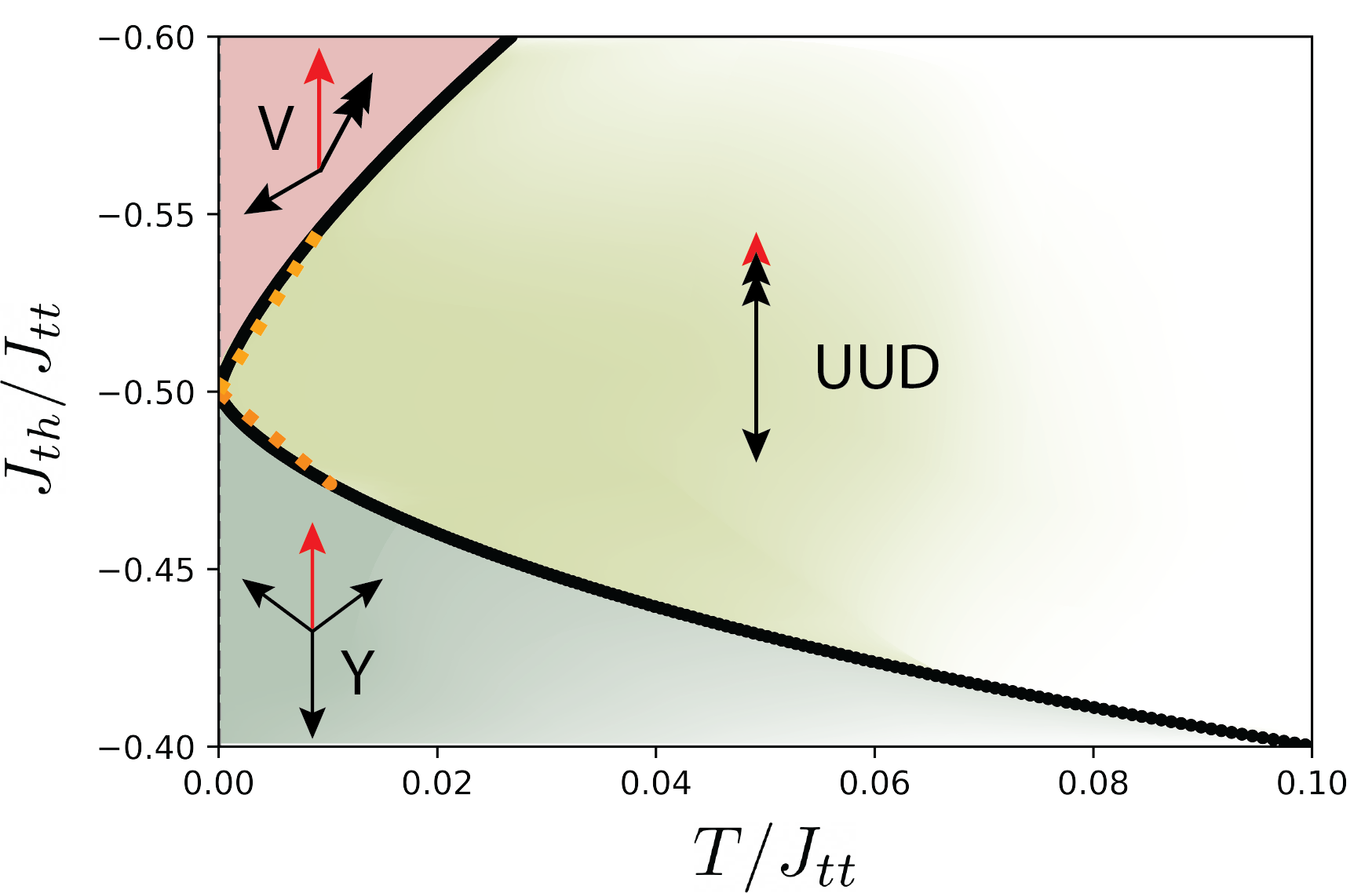}
\caption{Low temperature phase diagram of model~\eqref{Eq:Hamiltonian} around $J_{th}/J_{tt} = -1/2$ at fixed $K_{th} = 0$, $J_{hh} = -J_{tt} = -1$. The UUD phase emerges at finite $T$ from the value $J_{th}/J_{tt} = -1/2$ since it exhibits the maximal entropy reduction (order-by-disorder), as shown in Fig.~\ref{Fig:temp_fluct}(b). The phase boundaries are calculated by considering Gaussian spatial fluctuations of the three-sublattice ansatz with three triangular ($A, B, C$) and two honeycomb ($\alpha, \beta$) sublattices of Heisenberg spins, as described in the text. The dotted yellow line is the linearized phase boundary used to determine the effective $K_{th}^*$ in Eq.~\eqref{Eq:effective_Kth_thermal_sup}.}
\label{Fig:Window_UUD} 
\end{figure}

We numerically compare $F^{\mu}\left(j_{th},T\right)$ with $F^{\text{UUD}}\left(j_{th}^{\left(0\right)},T\right)$
to find the boundaries at finite temperatures for which the UUD state is favored. In Fig.~\ref{Fig:temp_fluct}(c) we show the free energy difference between UUD and the coplanar Y state at $j_{th}=j_{th}^{\left(0\right)}+0.03$. Due to the entropic term, the UUD state has a lower free energy for $T/J_{tt} > 0.012$. By performing this analysis for $j_{th}$ around $j_{th}^{(0)}$, we obtain the low temperature phase diagram shown in Fig.~\ref{Fig:Window_UUD} that hosts an UUD phase over a finite region close to $j_{th}^{(0)}$ at $T>0$.

Linearizing the slopes of the phase boundary between the UUD and the Y and V phases in
Fig.~\ref{Fig:Window_UUD}, we estimate these slopes 
to be $J_{th}/T \approx 2.5$ and $J_{th}/T \approx -5$, respectively.
The width of the UUD phase is thus approximately given by 
\begin{align}
\Delta J_{th} \approx 7.5T \,.
\end{align}
Comparing with the opening of UUD window the classical GS phase diagram in Fig.~\ref{Fig:PD}(a) and using the analytical expressions for the phase boundaries in Eqs.~\eqref{boundary1} and~\eqref{boundary2}, we obtain an effective biquadratic coupling of
\begin{align}
\frac{K_{th}^{*}}{J_{tt}}\approx-0.9\frac{T}{J_{tt}}\,.
\label{Eq:effective_Kth_thermal_sup}
\end{align}
For $ T=0.05 J_{tt}$ we estimate a thermally induced $K_{th}^* \approx -0.045 J_{tt}$ that is comparable to the effective $K_{th}^* \approx -0.04 J_{tt}$ induced by quantum fluctuations at $T=0$ for $S=1/2$. This puts the model in the vicinity of the Heisenberg limit in the phase diagram of Fig.~\ref{Fig:PD}(a). 

\section{Conclusions}
\label{Sec:Conclusions}
To conclude, we here design and study bilinear-biquadratic SO(3) symmetric Heisenberg models on stacked lattices that exhibit emergent three-state Potts-nematic order. The lattice designs consist of AFM coupled spins on a triangular sublattice that are interacting with spins on a FM coupled sixfold symmetric sublattice. We presented different lattice variants realizing this motif such as the windmill model (honeycomb-triangular bilayer), a triangular-kagome bilayer and an ABC stacked triangular lattice trilayer. 
We demonstrated that these models can be regarded as SO(3) invariant generalizations of Heisenberg and Ising triangular AFMs in magnetic fields, and largely extend the material space to study emergent Potts-nematic phases. We show that an effective biquadratic exchange coupling, which is explicitly included in our model as parameter  $K_{th}$, arises from quantum and thermal fluctuations already in a purely bilinear model (with $K_{th} = 0)$ via an order-by-disorder mechanism. This reduces the requirements for an experimental realization of the proposed stacked spin models. We thus hope that our work leads to further exploration of materials that realize SO(3) invariant spin exchange interactions on such stacked lattices. 

We here focus on the example of classical spin models on the 2D windmill lattice, and show that while magnetic order is absent at finite temperatures due to the Hohenberg-Mermin-Wagner theorem, the system hosts a finite temperature phase transition into a state with long-range three-state Potts-nematic order above the $T=0$ UUD ground state. Our key result is the phase diagram of this model at finite-T as a function of Hamiltonian parameters $J_{th}/J_{tt}$, $K_{th}/J_{tt}$, and $J_{hh}/J_{tt}$. We relate our findings obtained using large-scale MC simulations and analytical theory to known results on the Ising and Heisenberg triangular AFMs in field. For example, tuning $J_{th}/J_{tt}$ in our models is similar to tuning the magnetic field strength in the other models and changing $K_{th}/J_{tt}$ effectively changes the ``Isingness" or degree of collinearity of our model.  
This allows to smoothly tune between the Heisenberg and Ising limits, which both exhibit $\mathbb{Z}_3$ Potts transitions at finite temperatures (this holds both for our models and the Heisenberg and Ising triangular AFMs in an external magnetic field). 

We also introduce a tuning parameter to manipulate the Potts order that corresponds to the spatial inhomogeneity of the magnetic field (or spatial rigidity of the field) that the triangular spins experience. This rigidity is related to the spin correlation length on the honeycomb sublattice $\xi_{h}$ and is thus controlled by the tuning parameter $J_{hh}/J_{tt}$. 
Within our MC simulations we show that Potts $T_c$ suddenly drops to zero at a critical, small value of $J_{hh}/J_{tt}$. In analogy to the Ising model in a magnetic field~\cite{kinzelPhenomenologicalScalingApproach1981,qianCriticalFrontierTriangular2004}, we conjecture that this transition lies in the KT universality class. Our data is also consistent with a first-order transition and this question deserves further study. 

While we focused on the study of the $Z_3$ Potts-nematic phase transition above the UUD ground state, the zero-temperature phase diagram hosts states that have an additional $U(1)$ symmetry in their order parameter manifold. At finite temperature above those ground states, one expects the appearance of an additional KT transition, and the interplay and order in the sequence of KT and Potts phase transitions could be explored in the future. A study of the finite temperature phase diagram above the non-collinear umbrella ground state that occupies the $K_{th} > 0$ part of the $T=0$ phase diagram is also worthwhile. 

Another open question is the effect of spatial site or bond disorder, which is known to be able to induce non-collinear spin orderings. Finally, including additional degrees of freedom in the model such as phonons that couple to the Potts-nematic degree of freedom or considering a quantum spin version of the proposed models, e.g. with spin $S=1$ are interesting future research directions; for interesting studies of related quantum $S=1$ models on triangular and honeycomb lattices, see e.g.~\cite{Wang_Starykh_Chubukov_Batista-PRB-2017, Strockoz_Venderbos-arXiv-2022}.

\section*{Acknowledgements}
We thank P.~Chandra, P.~Coleman, R.~M.~Fernandes, R.~Flint, M.~Kornja\v ca, and J.~Schmalian for useful discussions. This research was supported by the U.S. Department of Energy, Office of Basic Energy Sciences, Division of Materials Sciences and Engineering. Ames National Laboratory is operated for the U.S. Department of Energy by Iowa State University under Contract No.~DE-AC02-07CH11358. V.L.Q. acknowledges support from the Research Corporation for Science Advancement via P.P.O.'s Cottrell Scholar Award.

\bibliographystyle{apsrev4-2}
\bibliography{Submission_PottsZ3.bib}

\clearpage

\appendix
\section{Variational ansatz and analytical solutions}
\label{SM_var_ansatz}

\begin{figure}[h!]
\centering
\includegraphics[width=0.95\columnwidth]{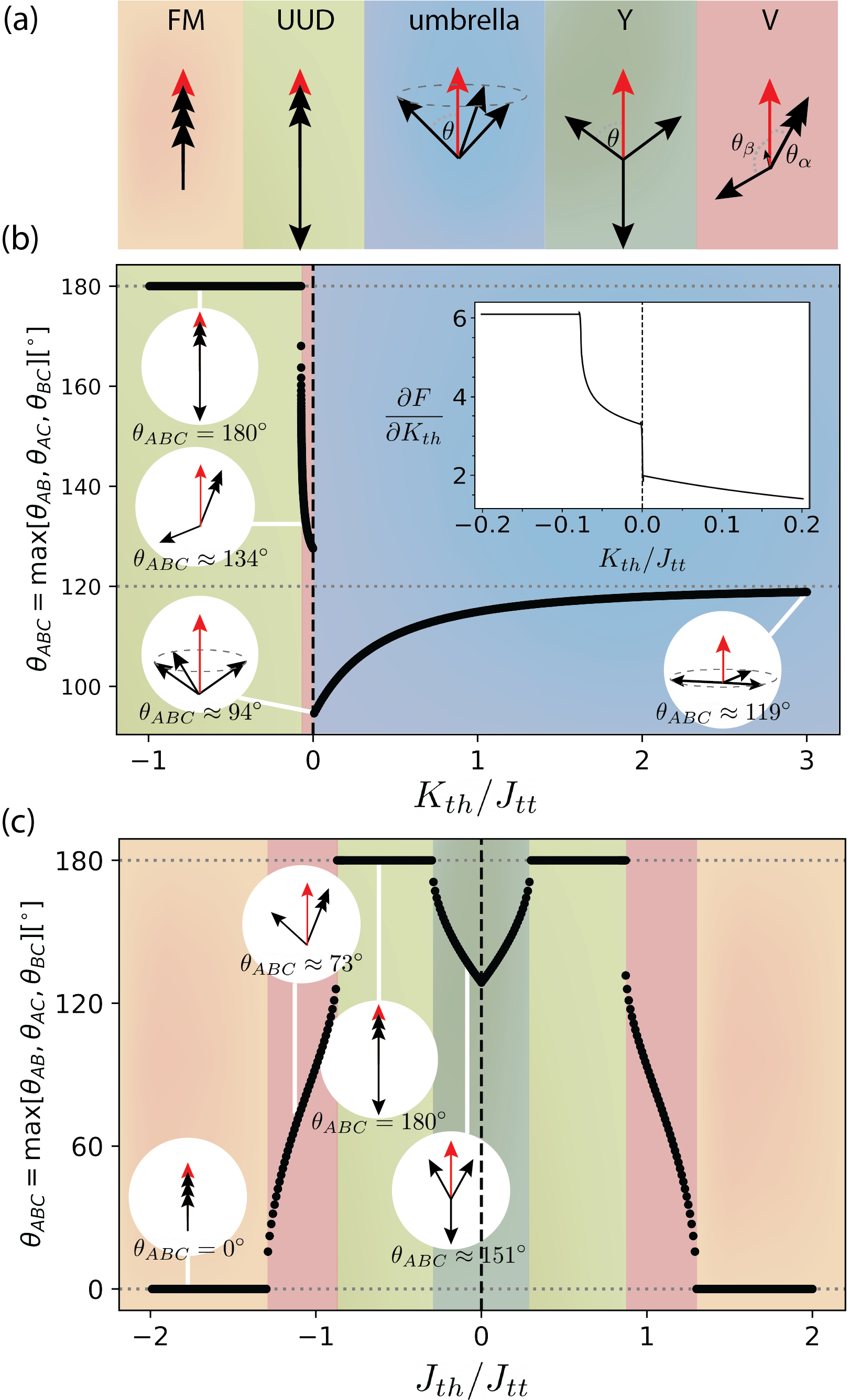}
\caption{ (a) Classical ground state phases for the three triangular sublattices (black) and the honeycomb sublattice (red) shown with the angles chosen to parametrize the states for the analytical expressions for the energies given in Eq.~\eqref{E_analyt}. (b,c) One-dimensional parameter cuts through the variational classical phase diagram in Fig.~\ref{Fig:PD}(a). Panel (b) is for fixed $J_{th}/J_{tt}=-0.8$ and panel (c) is for fixed $K_{th}/J_{tt}=-0.1$. The plots show the largest relative angle between any pair of triangular spins $\theta_{ABC}=\max{[\theta_{AB}, \theta_{AC}, \theta_{BC}]}$, which uniquely characterizes the different phases. Dashed black lines indicate first-order phase transitions within the variational ansatz. The inset in (b) shows the derivative of the energy with respect to $K_{th}$, indicating a continuous transition from the UUD to the V phase and a first-order transition from V to the umbrella phase.}
\label{Fig:cuts}
\end{figure}
Here we describe details of the variational calculation resulting in the zero-temperature classical phase diagram shown in Fig.~\ref{Fig:PD}(a). We introduce the variational ansatz, state the energies for all the phases that appear on the phase diagram, and analytically solve for the linearized boundaries between phases.

We consider the limit $|J_{hh}| \gg J_{tt}, |J_{th}|, |K_{th}|$ and introduce a variational ansatz assuming three triangular sublattices ($A, B, C$) and a honeycomb sublattice ($h$) as illustrated in the upper right corner of Fig.~\ref{Fig:PD}(a) of the main text. Specifically, we assume that the GS spin configuration is of the following form
\begin{align}
\begin{split}
    \boldsymbol{S}_A &= S(\sin \theta_A \cos \phi_A, \sin \theta_A \sin \phi_A, \cos \theta_A) \\
    \boldsymbol{S}_B &= S(\sin \theta_B \cos \phi_B, \sin \theta_B \sin \phi_B, \cos \theta_B) \\
    \boldsymbol{S}_C &= S(\sin \theta_C \cos \phi_C, \sin \theta_C \sin \phi_C, \cos \theta_C) \\
    \boldsymbol{S}_h &= S(0,0,1) \,,
\end{split}
\label{eq:var_ansatz}
\end{align}

for $S=1$. Note that spins on a given sublattice ($A, B, C, h$) point along the same direction in the complete system. Without loss of generality, we set the $z$ axis to point along the honeycomb spins.

The variational ansatz in Eqs.~\eqref{eq:var_ansatz} is exact in the limit $J_{hh} \to -\infty$ and $K_{th}=0$ when the model~\eqref{Eq:Hamiltonian} maps to the triangular Heisenberg antiferromagnet in an external magnetic field~\cite{Chubukov1991, Starykh2015}, because the honeycomb spins are then ordered FM at $T=0$. We have explicitly checked using MC simulations that this ansatz yields the correct ground states even in the regime of our main interest, which is for $|J_{hh}| \approx J_{tt}$ and $J_{th}, K_{th}<0$, $|J_{th}|, |K_{th}| < J_{tt}$.

We minimize the energy of the Hamiltonian given in Eq.~\eqref{Eq:Hamiltonian} of the main text with respect to the six free angles of the triangular spins in the local coordinate system associated with the orientation of the honeycomb spin.
To avoid being trapped in a local minimum, we repeat the local minimization procedure
20 times starting from random initial conditions.
We identify five classical ground state phases: FM, UUD, umbrella, Y, and V states, as illustrated in Fig.~\ref{Fig:cuts}(a). Next, we write the analytical expressions for the ground state variational energy per unit cell for each of the states:

\begin{align}
\begin{split}
\frac{E_{\rm{FM}}}{NS^2} &= 3J_{tt}+6J_{th}+6K_{th}+3J_{hh} \\
\frac{E_{\rm{UUD}}}{NS^2} &= -J_{tt} + 2J_{th}+6K_{th}+3J_{hh} \\
\frac{E_{\rm{umbrella}}[\theta]}{NS^2} &=\frac{3J_{tt}}{4}\left(1+3\cos{2\theta}\right) + 6J_{th} \cos{\theta} +\\ &+6K_{th}\cos^2{\theta} +3J_{hh}\\
\frac{E_{\rm{Y}}[\theta]}{NS^2} &=J_{tt}\left(\cos{2\theta}-2\cos{\theta}\right)+2J_{th}\left(-1+2\cos{\theta}\right) \\
&+2K_{th}\left(1+2\cos^2{\theta}\right) +3J_{hh}\\
\frac{E_{\rm{V}}[\theta_{\alpha}, \theta_{\beta}]}{NS^2} &= J_{tt}\left(1+2\cos{[\theta_{\alpha}+\theta_{\beta}]} \right) \\
&+2 J_{th}\left(2\cos{\theta_{\alpha}} + \cos{\theta_{\beta}}\right) + \\
&+2K_{th}\left(2\cos^2{\theta_{\alpha}} + \cos^2{\theta_{\beta}}\right)+3J_{hh},
\end{split}
\label{E_analyt}
\end{align}
where the angles $\theta$, $\theta_{\alpha}$ and $\theta_{\beta}$ used in the parametrization of the umbrella, Y and V states are chosen as illustrated in Fig.~\ref{Fig:cuts}(a). Note that the total magnetization on the triangular lattice in the V state denoted with the small black arrow is collinear with the honeycomb spin only for $K_{th}=0$.

From the minimization of the energy with respect to the angles, we obtain the analytic expression for the ground state energies for the umbrella and the Y states
\begin{align}
\begin{split}
\frac{E_{\rm{umbrella}}}{NS^2} &= -\frac{3 J_{tt}}{2} \frac{3+4\frac{K_{th}}{J_{tt}}+4\big(\frac{J_{th}}{J_{tt}}\big)^2}{3+4 \frac{K_{th}}{J_{tt}}},
\\
\frac{E_{\rm{Y}}}{NS^2} &= J_{tt}\frac{-3-4\big(\frac{J_{th}}{J_{tt}}\big)^2-8\frac{J_{th}K_{th}}{J_{tt}^2}+8\big(\frac{J_{th}}{J_{tt}}\big)^2}{2(1+2\frac{K_{th}}{J_{tt}})} .
\end{split}
\end{align}
The expression for the V state that has to be minimized over two angles $\theta_{\alpha}$ and $\theta_{\beta}$ cannot be obtained analytically.

The phase diagram obtained by the variational ansatz is shown in Fig.~\ref{Fig:PD}(a) of the main text. Fig.~\ref{Fig:cuts}(b) shows the reorientation of the spins as a function of biquadratic coupling $K_{th}$ for fixed $J_{th}/J_{tt}=-0.8$. Fig.~\ref{Fig:cuts}(c) is the corresponding curve as a function of $J_{th}$ for fixed $K_{th}/J_{tt}=-0.1$. The vertical axis illustrates the maximal relative angle between the triangular spins $\theta_{ABC}=\max{[\theta_{AB}, \theta_{AC}, \theta_{BC}]}$. In Fig.~\ref{Fig:cuts}(b), we show how the UUD state continuously reorients to the V state. At $K_{th}\rightarrow 0$, the V state reorients to the non-coplanar umbrella state through a first-order phase transition, as seen from the first derivative of the energy with respect to $K_{th}$ in the inset of Fig.~\ref{Fig:cuts}~(b). For large and positive values of $K_{th}/J_{tt}$, the angle between spins of the triangular and honeycomb sublattices in the umbrella state reaches $90^{\circ}$ and the relative angle between the triangular spins approach $120^{\circ}$.

Fig.~\ref{Fig:cuts}(c) 
shows that the model exhibits a continuous phase transitions from FM to V and from UUD to Y phases as the ratio $|J_{th}/J_{tt}|$ is decreased. In contrast, at larger negative $K_{th}$ the  transition between the UUD and the FM phase is first-order within the three-sublattice ansatz (not shown). 

From the analytical expressions for the energies of the variationally obtained states, 
we obtain the analytic solutions for the phase boundaries. These are stated in Eqs.~\eqref{boundary1} and~\eqref{boundary2}. The phase boundaries between two neighboring phases are described by (notice that the $y$-axis is reversed in Fig.~\ref{Fig:PD}(a)):
\begin{align}
\begin{split}
\mbox{FM and UUD} \qquad J_{th}/J_{tt} &= -1 \\
\mbox{FM and umbrella} \qquad J_{th}/J_{tt} &= -\frac{3}{2}-2K_{th}/J_{tt} \\
\mbox{UUD and Y} \qquad J_{th}/J_{tt} &= -\frac{1}{2}-2K_{th}/J_{tt}. \,
\end{split}
\label{boundary1}
\end{align}
The boundaries with the V phase cannot be determined analytically in the most general case when the relation between $\theta_{\alpha}$ and $\theta_{\beta}$ is unknown. In the limit $K_{th}\to 0^-$, however, expanding to second order around $\theta_{\alpha}=\theta_{\beta}=0$ for FM ($\theta_{\alpha}=0, \theta_{\beta}=\pi$ for UUD) and evaluating at $\theta_{\alpha}=\epsilon$ and $\theta_{\beta}=2\epsilon$ for FM ($\theta_{\alpha}=\epsilon$ and $\theta_{\beta}=\pi-2\epsilon$ for UUD) in the limit $\epsilon \to 0$, gives the linear boundaries with the V phase
\begin{align}
\begin{split}
\mbox{FM and V} \qquad J_{th}/J_{tt} &= -\frac{3}{2}-2K_{th}/J_{tt}\,, \qquad K_{th} \to 0^-\\
\mbox{UUD and V} \qquad J_{th}/J_{tt} &= -\frac{1}{2}+6K_{th}/J_{tt}\,, \qquad K_{th} \to 0^-.
\label{boundary2}
\end{split}
\end{align}

\section{Three-state Potts-nematic transition}
Here we show additional MC results regarding the $\mathbb{Z}_3$ Potts transition between the paramagnetic phase and the phase with long-range Potts-nematic order.
\subsection{Derivation of Binder cumulant for Potts order}
\label{SM_Binder_cumulant}

Here we derive the Binder cumulant~\cite{Binder1981, Binder1984} for complex $\mathbb{Z}_3$ Potts order parameter $\boldsymbol{m}_3$, introduced in Eq.~\eqref{Eq:order_parameter} of the main text.
We calculate the Binder ratio
\begin{align}
R_2 = \frac{\langle |\boldsymbol{m}_3|^4 \rangle}{\langle |\boldsymbol{m}_3|^2\rangle^2},
\label{R2_Binder}
\end{align}
where the averages are taken over the order parameter distribution $P(\boldsymbol{m}_3)$. In the ordered phase, the order parameter acquires a finite value $|\boldsymbol{m}_3|=1$ leading to $R_2=1$. 
In the disordered phase $|\boldsymbol{m}_3|$=0, so $R_2$ has to be calculated up to the second order around the mean value, including the fluctuations in order parameter.
The fluctuations of the order parameter $\boldsymbol{m}_3$ are Gaussian and lie in the two-dimensional complex plane. 
In the disordered phase, the fluctuations of the amplitude follow a normal distribution. In polar coordinates centered at zero, the distribution around the mean value is
\begin{align}
P_{\rho}(\rho) =\frac{1}{\sqrt{2\pi\sigma^{2}}}e^{-\frac{\rho^{2}}{2\sigma^{2}}},
\end{align}
while the angular contribution is uniformly distributed, $P_{\theta}(\theta)=1/(2\pi)$. Integrating over the whole space, we find $R_2=2$ in the disordered phase. 
 
Combining the previous results, we construct the Binder cumulant $U_2$ that gives $U_2=1$ in the ordered UUD and $U_2=0$ in the disordered phase,
\begin{align}
U_2=2\left(1-\frac{1}{2}R_2\right) \,,
\end{align}
identical with the Binder cumulant for the 2-component order parameter for the XY model \cite{Sandvik2010}.

The Binder ratio and the derived $\mathbb{Z}_3$ Potts Binder cumulant $U_2$ can be used as a precise measure of the $\mathbb{Z}_3$ Potts transition for which the leading finite-size corrections cancel out. The transition temperature is determined from the intersections of calculated Binder cumulants for different system sizes, as shown in the Fig.~\ref{Fig:order_parameter}(b) of the main text.

\subsection{Amplitude of Potts order parameter versus temperature}
\label{SM_order_param}

\begin{figure}[t]
    \centering
    \includegraphics[width=0.95\columnwidth]{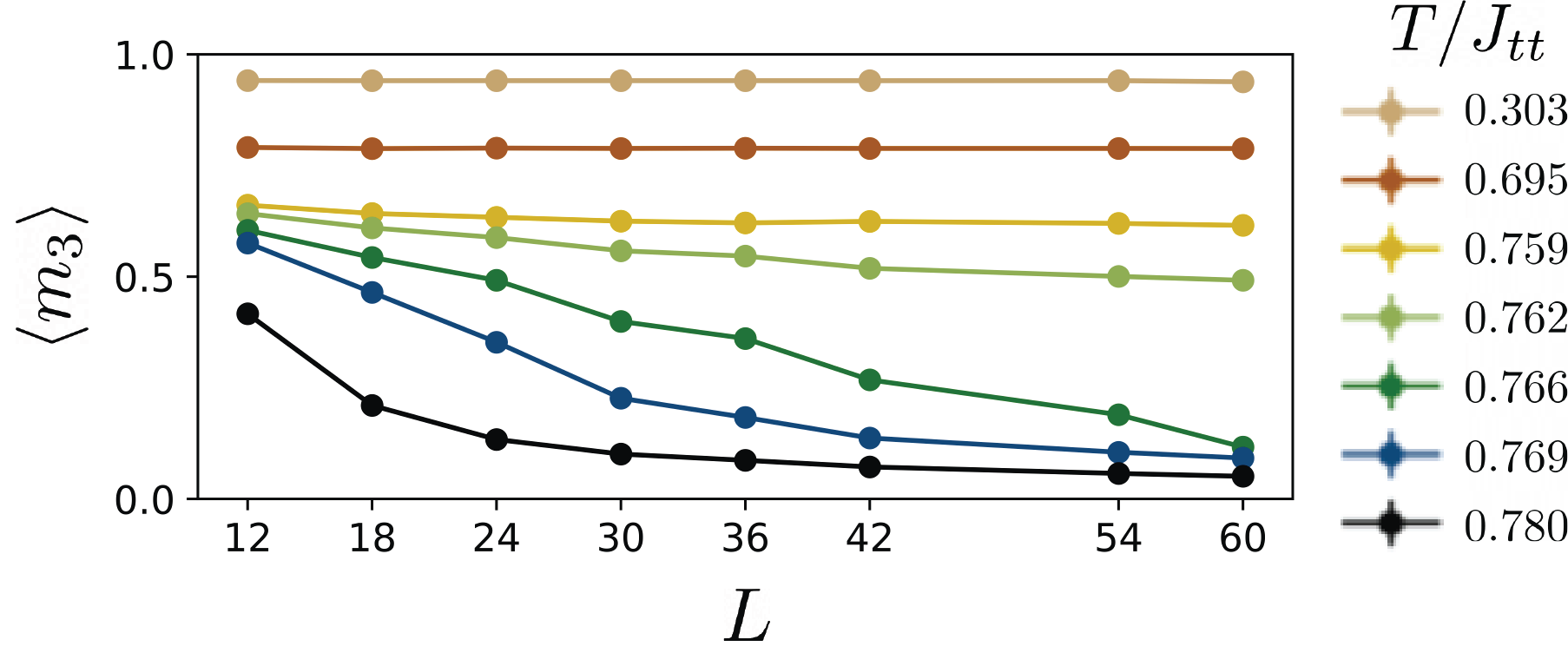}
    \caption{The amplitude $\langle m_3 \rangle$ of the order parameter as a function of linear lattice size $L$ shows a saturation for $T<T_c$ and an exponential decay for $T>T_c$.}
    \label{Fig:m3_vs_T}
\end{figure}

As a complementary indicator of $\mathbb{Z}_3$ Potts transition to the results shown in Fig.~\ref{Fig:order_parameter} in the main text, we show the $\mathbb{Z}_3$ Potts magnetization $\langle m_3\rangle$ as a function of system size in Fig.~\ref{Fig:m3_vs_T}. The change in the trend in the decay of $\langle m_3\rangle$ as a function of linear system size $L$ from saturation for $T<T_c$,
to exponential decay for $T>T_c$ indicates the $\mathbb{Z}_3$ Potts transition.

\begin{figure*}[t]
    \includegraphics[width=\linewidth]{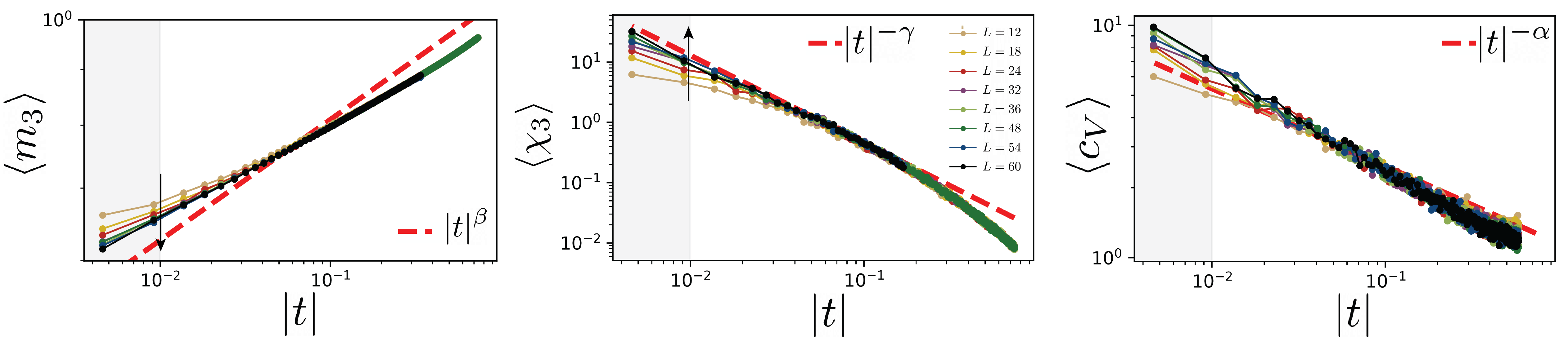}
    \caption{$\mathbb{Z}_3$ Potts magnetization $\langle m_3 \rangle$, $\mathbb{Z}_3$ Potts susceptibility $\langle \chi_3 \rangle$ and specific heat $\langle c_V \rangle$ obtained in our calculations are compared with the analytical exponents expected for this transition as a function of reduced temperature $|t|$ for $T<T_c$. The Monte-Carlo results for different system sizes $L$ are shown on log-log scale compared with the critical power-law behavior for the $\mathbb{Z}_3$ Potts model (dashed red lines). The region close to the critical point is shaded. The black arrows indicate the trends in scaling functions with the increased system sizes.}
    \label{Fig:scaling}
\end{figure*}

\subsection{Scaling behavior of raw Monte-Carlo results as a function of reduced temperature}
\label{SM_scaling_analysis}

In Fig.~\ref{Fig:scaling_collapse}, we show a finite size scaling analysis of Potts magnetization $m_3$, Potts susceptibility $\chi_3$ and specific heat $c_V$ using the theoretically known exponents of the 2D $\mathbb{Z}_3$ Potts universality class: $\alpha = 1/3$, $\beta=1/9$ and $\gamma = 13/9$ \cite{Wu1982}. In Fig.~\ref{Fig:scaling}, we plot these observables on a double logarithmic scale as a function of the reduced temperature $t=\frac{T-T_c}{T_c}$ and compare to the slope obtained from the power law using the exact Potts exponents. Here, $T_c$ is extracted from the crossing of the Binder cumulant and slightly tuned within the determined error bar from optimization of the scaling collapse given in Fig.~\ref{Fig:scaling_collapse}. The results shown in Fig.~\ref{Fig:scaling} show good agreement with the power laws of the exact Potts critical exponents. The deviations from linearity get smaller for larger system sizes and the agreement with the predictions using the exact critical exponents also improves with larger system size, as indicated by the black arrows. 

\section{Expressions of matrix for Gaussian thermal fluctuations of
coplanar states}
\label{SM_thermal_fluc_expressions}

This section contains the explicit expressions of the effective Hamiltonians $\mathcal{H}^\mu$ that correspond to coplanar ground states. As described in the Section~\ref{SM_thermal_fluc}, $\mathcal{H}^\mu$ includes Gaussian thermal fluctuations around the zero temperature configurations. We expand to second order in the fluctuations of the angles $(\delta\theta_{a,i},\delta x_{a,i})$
where $\delta x_{a,i}=\sin{\theta_{a,i}}\delta\phi_{a,i}$ around a given ground state configuration $\theta_{a}^{0}$ and $\phi_a^{0}$. Here, $i$ labels the spatial index and $a\in\{A,B,C,\alpha,\beta\}$ is the sublattice index of the three-sublattice variational ansatz we consider for the ground states. The angles $\theta$ and $\phi$ describe polar and azimuth with respect to the honeycomb spin at basis site $a$. Interestingly, we find that the fluctuations in $\delta \theta_{a,i}$ and $\delta x_{a,i}$ decouple for coplanar states and the effective Hamiltonian for coplanar ground states thus takes the form  
\begin{align}
\delta\mathcal{H}=\delta\mathcal{H_{\delta\theta}}+\delta\mathcal{H}_{\delta x}.\label{Eq:sectors}
\end{align}
We further find that $\delta\mathcal{H}_{\delta x}$ is independent of the set of ground state angles $\{\theta_a^0, \phi_a^0\}$. Which states are preferred by the entropy term
is thus solely determined by $\delta\mathcal{H}_{\delta\theta}$. Similar findings about the separation of the angular fluctuations and the insignificance of  $\delta\mathcal{H}_{\delta x}$ in the order-by-disorder mechanism was reported for the Heisenberg AFM on the triangular lattice in a magnetic field~\cite{Kawamura1984}.

Due to the decoupling of the angular fluctuations, the matrix $\hat{A}_{\mathbf{q}}$ takes the following form 
\begin{align}
\hat{A}_{\mathbf{q}}=\begin{pmatrix}\hat{M}_{\mathbf{q}} & \hat{0}\\
\hat{0} & \hat{N}_{\mathbf{q}}
\end{pmatrix}\,\,.
\end{align}
The diagonal block describing the coupling between the $\delta \theta_{a,i}$ variables is given by
\begin{align}
\hat{M}_{\mathbf{q}}=\begin{pmatrix}3J_{tt} & uJ_{tt}\varepsilon_{\mathbf{q}} & wJ_{tt}\varepsilon_{\mathbf{q}}^{*} & \cos{\theta_{A}^{0}}J_{th}\xi_{\mathbf{q}} & \cos{\theta_{A}^{0}}J_{th}\xi_{\mathbf{q}}^{*}\\
 & 3J_{tt} & vJ_{tt}\varepsilon_{\mathbf{q}} & \cos{\theta_{B}^{0}}J_{th}\xi_{\mathbf{q}} & \cos{\theta_{B}^{0}}J_{th}\xi_{\mathbf{q}}^{*}\\
 &  & 3J_{tt} & \cos{\theta_{C}^{0}}J_{th}\xi_{\mathbf{q}} & \cos{\theta_{C}^{0}}J_{th}\xi_{\mathbf{q}}^{*}\\
 & \mbox{c.c.} &  & 6\frac{J_{th}^{2}}{J_{tt}}-3J_{hh} & J_{hh}\xi_{\mathbf{q}}\\
 &  &  &  & 6\frac{J_{th}^{2}}{J_{tt}}-3J_{hh}
\end{pmatrix}
\end{align}
where $\theta_{A}^{0}$, $\theta_{B}^{0}$ and $\theta_{C}^{0}$ are the angles in the ground state, measured with the respect of the axis of the honeycomb
spin $\alpha$. For example, for the UUD state we find $(\theta_{A}^{0}$, $\theta_{B}^{0}, \theta_{C}^{0} = (0,0,\pi)$. We have also introduced the variables $u=\cos{(\theta_{A}^{0}-\theta_{B}^{0})}$, $v=\cos{(\theta_{B}^{0}-\theta_{C}^{0})}$,
$w=\cos{(\theta_{A}^{0}-\theta_{C}^{0})}$.
The block describing the coupling between the $\delta x_{a,i}$ variables is given by
\begin{align}
\hat{N}_{\mathbf{q}}=\begin{pmatrix}3J_{tt} & J_{tt}\varepsilon_{\mathbf{q}} & J_{tt}\varepsilon_{\mathbf{q}}^{*} & J_{th}\xi_{\mathbf{q}} & J_{th}\xi_{\mathbf{q}}^{*}\\
 & 3J_{tt} & J_{tt}\varepsilon_{\mathbf{q}} & J_{th}\xi_{\mathbf{q}} & J_{th}\xi_{\mathbf{q}}^{*}\\
 &  & 3J_{tt} & J_{th}\xi_{\mathbf{q}} & J_{th}\xi_{\mathbf{q}}^{*}\\
 & \mbox{c.c.} &  & 6\frac{J_{th}^{2}}{J_{tt}}-3J_{hh} & J_{hh}\xi_{\mathbf{q}}\\
 &  &  &  & 6\frac{J_{th}^{2}}{J_{tt}}-3J_{hh}
\end{pmatrix}
\end{align}
Both blocks contain the functions
\begin{align}
\varepsilon_{\mathbf{q}}=e^{iq_{x}a}+e^{\frac{-iq_{x}a+i\sqrt{3}q_{y}a}{2}}+e^{-\frac{iq_{x}a+i\sqrt{3}q_{y}a}{2}}
\end{align}
\begin{align}
\xi_{\mathbf{q}}=e^{-\frac{i\sqrt{3}q_{y}a}{3}}+e^{\left(\frac{iq_{x}a}{2}+\frac{i\sqrt{3}q_{y}a}{6}\right)}+e^{-\frac{iq_{x}a}{2}+\frac{i\sqrt{3}q_{y}a}{6}}
\end{align}
where $a$ is the lattice size of the triangular sublattice.

\section{Mapping to the Ising limit }
\label{Appendix:Mapping_Ising}

In this Appendix, we show how the model of Eq.~\eqref{Eq:Hamiltonian} maps to the Ising triangular AFM in a magnetic field at finite temperatures and $K_{th} < 0$, in the limit $|J_{hh}|, |K_{th}| \gg J_{tt},|J_{th}|$. We are going to do so by analyzing what each coupling constant is enforcing, respecting the hierarchy of the energy scales. 


First, the large $|J_{hh}|$ reduces the fluctuations of the honeycomb spins. While the honeycomb lattice does not order at finite temperatures, the large coupling allows us to write  
\begin{align}
    \mathbf{S}_{hi}=\left(\sin\theta_{i},0,\cos\theta_{i}\right).
\end{align}
with a slowly varying angle $\theta_{i}$, i.e., $|J_{hh}| \gg T$. Now, we address how the triangular lattice reacts to this texture of the honeycomb spins. For large $K_{th}<0$, the orientation of a given triangular spin $\mathbf{S}_{ti}$ is enforced to be collinear with the 
(mean-field) average local field of the nearest-neighbor honeycomb spins
\begin{align}
\mathbf{S}_{ti}=\frac{\sigma_i}{6} \sum_{j'=1}^{6} \mathbf{S}_{hj'}, \quad \sigma_i = \pm 1 \,. \label{eq:Sti_MF}
\end{align}


Rewriting the terms with $J_{tt}$ and $J_{th}$ of the Hamiltonian Eq.~\eqref{Eq:Hamiltonian} using Eq.~\eqref{eq:Sti_MF} and keeping only terms up to order $\delta \theta_{ij}=\theta_{i}-\theta_{j}$ (with $i,j$ nearest neighbors) and neglecting constant terms, one obtains an effective Ising model on the triangular lattice in a magnetic field,
\begin{align}
\mathcal{H}_{\text{eff}} = J_{tt}\sum_{\substack{\langle i,j\rangle_{tt}}
}\sigma_i \sigma_j + 6 J_{th} \sum_i \sigma_i.
\label{Eq:mapping_Ising}
\end{align}
This is the effective Ising model on the triangular lattice discussed in the main text.

\end{document}